\documentclass[a4paper,12pt]{amsart}
\usepackage{amssymb,amsmath,mathrsfs}
\usepackage[usenames,dvipsnames]{xcolor}
\usepackage{hyperref}
\usepackage{tensor}
\usepackage{graphicx}
\usepackage{revsymb}
\usepackage[left=1.2in,top=1in,right=1.2in,bottom=1in,headheight=0.8in,foot=0.5in]{geometry}
\usepackage[nodisplayskipstretch]{setspace} 
\usepackage{mdframed}
\usepackage{tikz}

\hypersetup{
	colorlinks=true,         
	linkcolor=MidnightBlue,          
	citecolor=MidnightBlue,
	urlcolor=MidnightBlue            
 }
\usepackage{natbib}
\setcitestyle{aysep={}} % author date

\newcounter{wn}
\newcommand{\BOstep}{\noindent\textbf{Step \stepcounter{wn}\arabic{wn}:} }

\title{\sc Finding Time for Wheeler-DeWitt Cosmology}
\date{Draft of \today.}

\usepackage{amsaddr}

\author{Nick Huggett}
\address{\vspace{-0.8pc}University of Illinois Chicago \\Chicago, Illinois, USA}
\email{\href{huggett@uic.edu}{huggett@uic.edu}}

\author{Karim P. Y. Th\'ebault}
\address{\vspace{-0.8pc}University of Bristol\\Bristol, United Kingdom}
\email{\href{mailto:karim.thebault@bristol.ac.uk}{karim.thebault@bristol.ac.uk}}

\makeatletter
\let\uppercasenonmath\@gobble

\begin{document}
\setstretch{1.2}
\maketitle

\begin{abstract}
We conduct a case study analysis of a proposal for the emergence of time based upon the approximate derivation of three grades of temporal structure within an explicit quantum cosmological model which obeys a Wheeler-DeWitt type equation without an extrinsic time parameter. Our main focus will be issues regarding the consistency of the approximations and derivations in question. Our conclusion is that the model provides a self-consistent account of the emergence of chronordinal, chronometric and chronodirected structure. Residual concerns relate to explanatory rather than consistency considerations. 
\end{abstract}
\tableofcontents
\setstretch{1.4}
\section{Introduction}

\textit{Time may change me, but I can't trace time}. An immortal philosopher once said. But could the \textit{trace of time} itself change? Might the \textit{structure of time} be \textit{emergent}? And, if so, could such temporal structure not \textit{de}-emerge? And then even \textit{re}-emerge again? In this paper we will consider, in detail, the formal and physical features of a quantum cosmological model due to \cite{Kiefer:1988} and \cite{kiefer:1995} which displays precisely such remarkable emergent temporal structure.    

By temporal structure we will mean the basic features of time that provide us with a temporal betweenness relation, or chron\textit{ordinal} structure, a quantitative duration relation between events, or chrono\textit{metric} structure, and temporal directionality, or chrono\textit{directed} structure. By emergent we will mean derivation under some approximation. 

Our central goal will be to explore, in the context of a concrete model, a cluster of challenges to the temporal emergentist story that each seek to undermine the relevant approximate derivations on the grounds that they assume what they are aiming to show. We will consider a specific challenge in the context of the appeal to the Born-Oppenheimer (BO) and Wentzel-Kramers-Brillouin (WKB) approximations in the approximate derivation of chronordinal and chronometric structures due to \cite{chua:2021}, and the idea of `temporal double standards' due to \cite{price:1996} in the context derivation of chronodirected structure. 

The paper is organised as follows: \S\ref{EmergenceofTime} provides more details regarding what we mean by the structure and emergence of time. \S\ref{TSinQC} contains an extended discussion of explicit quantum cosmological models in context of the emergence of time.
In \S\ref{WDWordmetric} we consider the application of the Born-Oppenheimer approximation to a simple quantum cosmological model, in comparison with its standard application in molecular chemistry. In \S\ref{approxcocm} we then consider the approximate derivation of chronordinal and chronometric structure in the semi-classical approximation. Finally, in \S\ref{sec:approxCD} we examine the potential for failure of the semi-classical approximation to be remedied via a decoherence based argument that contains a mechanism for the potential emergence of chronodirected structure. \S\ref{TDS} considers sceptical arguments that challenge the cogency of the temporal emergentist story based upon the idea of temporal double standards, or assuming the temporal structure that you are trying to derive. In particular, we provide reconstructions of double standards arguments regarding the emergence of chronordinal and chronometric structure due to \cite{chua:2021} and  regarding the emergence of chronodirected structure due to \cite{price:1996}. In each case, the relevant charges is found to fail. The model thus provides a self-consistent account of the emergence of chronordinal, chronometric and chronodirected structure. Residual concerns related to explanatory rather than consistency considerations. 

\section{The Structure and Emergence of Time}
\label{EmergenceofTime}

Even restricting attention to mechanics, physical time is not a unitary concept; rather, we can distinguish various properties and structures. So doing will help us analyse the emergence of time in quantum cosmology later in the paper. They are partially ordered, so we will label them `grades' of temporal structure.\footnote{This account of the structure of time builds on \cite{Gryb:2024} which focuses on analysis of `chronordinal' and `chronometric' structure in the context of the so-called problem of time.  For a fascinating and wide-ranging philosophical investigation of temporal structure we highly recommend \cite{newton:1982}.} They should be thought of as structuring  spacetime in the first place, though here we will define them in terms of relations between `events' in a \emph{linear order}. In the case of Friedmann-Lemaitre-Robinson-Walker (FLRW) universes that concerns us here, these `events' are the well-defined instants of `cosmological time'. While in more general relativistic spacetimes, the corresponding relations can be defined as \textit{invariant structures} for points on \emph{timelike curves}, in terms of the causal and metrical structures of spacetime, or as \textit{surplus structures} based upon a space-like foliation \citep{Gryb:2024}.

The first grade of temporal structure that we consider orders sets of events with respect to a 3-place relation of temporal `betweeness': such a `chronordinal'  structure is that of an \textit{undirected temporal line}, which McTaggart named `C-series time'.\footnote{For more general discussion on the status of time ordering structure we strongly recommend the work of \cite{Farr:2012,farr:2016,farr2020c}. Building upon upon the ideas of \cite{mctaggart:1908}, \cite{reichenbach:1956} and \cite{black:1959}, Farr advocates a `C theory of time' in which there exists a undirected causal relation sufficient to define a partial ordering on the space of events.}  That is, given three chronordinally structured events, $a$, $b$, $c$, it is determinate, for example, that $b$ is \textit{temporally between} $a$ and $c$ but not (necessarily) whether it is \textit{before} or \textit{after} $a$. We assume that $a$ is not also between $a$ and $c$ (ruling out cyclic time) though of course one could consider relaxing that assumption.\footnote{Moreover, if $b$ is between $a$ and $c$, and $d$ is between $a$ and $b$, then $d$ is between $a$ and $c$. In relativistic spacetime these temporal lines can be specified invariantly as timelike curves, defined in terms of the lightcone structure; on such a curve, $b$ is temporally between $a$ and $c$, iff it lies in their `causal diamond'.}
    
The second grade of temporal structure that we shall consider is  \emph{chronometric}: the temporal distance relations between events. For every pair $\{a,b\}$ of chronordinally structured events there is also a unique (since we consider non-cyclic time) non-negative real number, the \textit{duration} of the interval between them, $\tau_{ab}$ (with $\tau_{ab}=0$ for $a=b$), satisfying the `triangle equality': for any triple of events $a$, $b$, $c$, the sum of two of the durations equals the third, $\tau_{ac}=\tau_{ab}+\tau_{bc}$, say. 

However, we can strengthen chronordinal structure in another way, by imposing \emph{chronodirected} structure on chronordinally structured events: the relation of `not-later-than' between pairs, $a\leq b$ ($b$ is between $a$ and $c$, if $c\leq a \wedge c\leq b \wedge b\leq a$, assuming that they are distinct). This relation gives McTaggart's B-series, our third grade of temporal structure.

Chronometric and chronodirected structures are logically independent. On the one hand, chronodirected structures are often not posited within physical theories with a temporal metric,  leading to a long tradition of attempts to derive an arrow of time from special initial conditions (e.g., \citet{sep-time-thermo}). On the other hand, one could envision a chronodirected world in which no durations are given; times organised as a directed line with no fact of the matter about the duration between pairs of points. However, if the set of events possesses a `clock', a real-valued quantity that is strictly increasing with respect to some total ordering of time, then it will provide both chronometric and chronodirected -- hence chronordinal -- structure: the duration between two events is the difference between their times, and $a$ is earlier than $b$ just in case $a$'s time is less than $b$'s. 

Now, when a set of objects possesses structures with the formal properties that we have described, it is still legitimate to ask whether they are in fact \emph{temporal}: perhaps instead they refer to some other physical relations (spatial relations, for instance). We take the general attitude that a structure can be identified as temporal in virtue of the roles it plays in dynamical physical theories. We do not offer a general account of what these roles are; rather we appeal to the fact that time is already identified in existing physical theories, and in this paper aim to pinpoint the (emergent) structures playing those roles.\footnote{See \citet{huggett2021nowhere} for more on such spacetime functionalism.}

This is not the only attitude one could take to the temporal nature of the structures. Suppose one posits that temporal structures are metaphysically `absolute'. So doing is compatible with our position, since we do not claim that temporal structures are identified \emph{with} their roles in physical theories, but \emph{by} them. However, one might deny even our weaker claim, so that temporal structures might come apart from physical theories. For instance, suppose a temporally oriented world, in which entropy is a monotonic function; even if one identifies the orientability of time by the arrow of increasing entropy, there would still remain the question of whether that arrow points to the absolute future or past. On our view, on the contrary, it is a matter of definition that the entropic arrow points to the `future', since that is how we use the term (assuming the alignment of that arrow with those of other physical processes). There is no further issue once we assume that temporal structures are identified by their roles in physical theory.\footnote{This attitude is compatible with asking how posited temporal structures that do not play a role in physical theory relate to those that do. For instance, does entropy increase or decrease with respect to growth of the `growing block'?} In short, once we have identified an emergent arrow that aligns with familiar temporal arrows, we are done; we do not need to -- and so won't! -- argue that it is future directed.

Having clarified what we will mean by temporal structure, let us next explain what we will mean by `emergence' of such a structure in a quantum cosmological model. We are interested in situations in which the following three features obtain:
\begin{itemize}
\item [1.]  A given temporal structure is not explicitly specified in the basic equations of a quantum cosmological model.
\item [2.] There exists a formal derivation of the relevant structure, or an approximation to such a structure, in some limit of the model. 
\item [3.] The formal derivation can be physically justified in low-energy and/or semi-classical regimes.
\end{itemize}

This approach is along the lines of the view of emergence, common in physics and becoming more popular in philosophy, in which emergence is compatible with reduction. We will assume, and not argue for, the basic structure of such an approach in what follows.\footnote{For more details see \cite{butterfield:2011,huggett2021nowhere, huggett2021spacetime,Palacios:2022}.}

Our focus in the paper will be a package of approaches to quantum cosmology, developed both separately and in collaboration by the physicists Claus Kiefer and Dieter Zeh, in which all three grades of temporal structure (putatively) emerge in our sense.\footnote{See in particular \cite{zeh:1986,Kiefer:1988,zeh:1989,kiefer:1995,Kiefer:2012}.}

\section{Emergence of Temporal Structure in Quantum Cosmology}
\label{TSinQC}
\bigskip
\begin{quote}
 ...the ``time ordering of events'' is a notion devoid of meaning [...] the concept of spacetime and time itself are not primary but secondary ideas in the structure of physical theory. These concepts are valid in the classical approximation. However, they have neither meaning nor application under circumstances when quantum-geometrodynamic effects become important. Then one has to forgo that view of nature in which every event, past, present, or future, occupies its preordained position in a grand catalog called ``spacetime''. There is no spacetime, there is no time, there is no before, there is no after. The question what happens ``next'' is without meaning. \newline
 \flushright \cite[p. 1124]{Wheeler:1968}
\end{quote}
\bigskip

The full gravitational Wheeler-DeWitt equation \citep{DeWitt:1967} is a heuristic semi-mathematical expression that results from informal application of the Dirac constraint quantization algorithm \citep{Dirac:1964} to the Hamiltonian formulation of general relativity \citep{Dirac:1958b,ADM:1960}. Famously, as indicated by Wheeler himself in the quote above, the equation does not contain any extrinsic temporal structure. Its formal and physical justification, both in general and in specific simple cosmological applications is open to dispute. We will not enter into these debates here.\footnote{For discussion see \cite{gryb:2011,gryb:2014,Gryb:2015,Gryb:2016a,Gryb:2017a,gryb:2018,Gryb:2024}. Note that a full understanding of Wheeler's assertion (which will not be necessary here) requires consideration of the `problem of time'.} Rather, we will assume the general pattern of argument leading to the Wheeler-DeWitt equation to be well enough justified and consider the explicit details of a particular model resulting from the quantization of symmetry reduced minisuperspace cosmology. 

Explicit model building in the context of quantum cosmology is usually based upon the quantization of classical FLRW cosmologies, in which spacetime can be foliated into homogeneous and isotropic spatial slices (or their generalization into Bianchi spacetimes, with homogeneous but anisotropic spatial slices). The only remaining variables are a scale factor $a$ for the spatial geometry, capturing the gravitational degrees of freedom, and the chosen matter degrees of freedom: for instance $\phi$ describing a spatially homogeneous scalar field. (For future reference, it is worth noting that these degrees of freedom can be thought of as mean field values for more realistic models, with other gravitation and field degrees of freedom appearing as higher moments that are ignored in a first approximation.) The space parameterized by these quantities is `mini-superspace' (MSS), and the standard form of the Wheeler-DeWitt equation in quantum cosmology (QC) is in the mini-superspace representation: an equation for a wavefunction $\Psi(a,\phi)$.\footnote{Analysis of these models goes back to \cite{DeWitt:1967} and includes notable work by \cite{blyth:1975,gotay:1980,hartle:1983,vilenkin:1984,kuchavr:1989,bojowald:2001,ashtekar:2006a}. See \cite{ashtekar:2021,gielen:2022} and citations therein for recent work. In what follows we will focus on the work on Claus Kiefer and provide full citations. Further work includes study of symmetry reduced but infinite dimensional midi-superspace models and so-called Gowdy models that feature gravitational waves. See for example, \cite{barbero:2010}, \cite[\S6]{ashtekar:2011}, \cite{tarrio:2013,deBlas:2017}.} This formulation brings the disappearance of time in quantum gravity into sharp relief: what kind of universe is described by a solution in which $\Psi(a,\phi)$ has support across a large swathe of MSS? It seems to describe amplitudes for all kinds of spatial slices, and no evolution through them (see \citet{warrier2022case} for a recent discussion of such issues) -- returning us to the quotation from Wheeler. 

\subsection{Wheeler-DeWitt meets Born-Oppenheimer}
 \label{WDWordmetric}
 
Even in such a restricted context as MSS finding solutions to the relevant equations requires semi-classical approximations; typically one or both of the WKB-approximation and the BO-approximation.\footnote{Early work is \cite{banks:1985,halliwell:1985,Kiefer:1988,vilenkin1989interpretation}. See \cite{kiefer:2005,kiefer:2013,kuchavr:2011,kiefer:2022,maniccia:2022} for   reviews.} Here we consider some of the explicit details of such an approach to the quantization of an MSS model with no cosmological constant, found in \cite{Kiefer:1988}. The Wheeler-DeWitt equation of the model is:
\begin{equation}
\label{WDWmini}
\hat H\Psi(\alpha,\phi)=\big{[} \frac{2}{3\pi m_p^2}\frac{\partial^2}{\partial\alpha^2} - \frac{\partial^2}{\partial \phi ^{2}} -\frac{3\pi m_p^2}{2}ke^{4\alpha} +m^2e^{6\alpha}\phi^2 \big{]}\Psi(\alpha,\phi)=0,
\end{equation}
where $k\in\{-1,0,+1\}$ is the sign of the spatial curvature, the scale factor $a$ (relative to some reference scale $a_0$ which we set to 1) is written in terms of a logarithmic variable $\alpha=\ln a$, the matter content is given entirely by a homogeneous scalar field $\phi$ with mass $m$, and $m_p\gg m$ is the Planck mass.\footnote{Note our treatment here, mainly following \cite{Kiefer:1988}, does not include full details of the quantum formalism. A fully rigourous presentation of this model would require explicit definition of the Hilbert space and operators and would encounter issues with self-adjointness related to the $a \rightarrow 0$ singularity. Neglect of these features is justified here since the behaviour we are interested in is far away from the singular `region'. See \cite{Thebault:2023}.} 

Our Wheeler-DeWitt equation has the form
\begin{equation}\label{eq:SETot} 
    (\hat T_1 +\hat T_2 +\hat W)\Psi(x_1,x_2)=E,
\end{equation}
with $E=0$. The Hamiltonian is the sum of the `kinetic energies' ($\sim\partial^2/\partial x_i^2$) of two subsystems, with a potential term that describes their interaction. Moreover, for a realistic field, the masses of the two subsystems are very different: the Planck mass of the gravitation field is much greater than that of the matter field, $m_p\gg m$. Thus the equation is formally analogous -- with $E=0$ -- to the time-independent Schr\"odinger equation for the kind of molecular system, containing heavy nuclei and light electrons, for which the Born-Oppenheimer (BO) approximation was developed \citep{bo:1927,born1955dynamical}. However, we will see that the rationale for its use in quantum cosmology is importantly different to that in atomic physics, a point that is relevant to addressing recent criticisms of derivations like Kiefer's. 

In the molecular case, the BO approximation can be understood as a three step method (see \citet{Jecko:2014} for a rigorous treatment.):\\

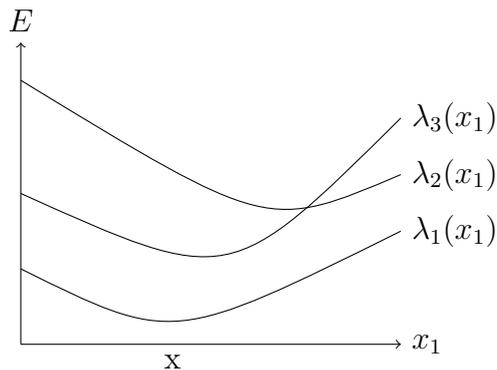
\begin{figure}
    \centering

\begin{tikzpicture}

\draw[->] (0,0) -- (2,0) node[anchor=north]{$\mathrm{x}$} -- (5,0) node[anchor=west]{$x_1$};
\draw[->] (0,0) -- (0,4) node[anchor=south]{$E$};
\draw (0,1) .. controls (2,0) .. (5,1.5) node[anchor=west]{$\lambda_1(x_1)$};
\draw (0,2) .. controls (2.75,.75) .. (5,3) node[anchor=west]{$\lambda_3(x_1)$};
\draw (0,3.5) .. controls (3.25,1.5) .. (5,2.25) node[anchor=west]{$\lambda_2(x_1)$};

\end{tikzpicture}

    \caption{The eigenvalues $\lambda_n$ of the reduced Hamiltonian, as (hypothetical) functions of the heavy degrees of freedom $x_1$. In the region around $x_1=\mathrm{x}$ the first three electronic energy levels can be seen to be widely separated: specifically, by far more than the kinetic energy of the nuclei. This is the condition for stable molecules, and for the BO approximation.}
    \label{fig:BO}

\end{figure}

\BOstep solve the (normalized) eigenvector equation for the `reduced Hamiltonian':

\begin{equation}
\label{eq:BO1}
    \big(\hat T_2+\hat W(x_1)\big) \psi_n(x_1;x_2) = \lambda_n(x_1) \psi_n(x_1;x_2).
\end{equation}
This equation treats the `heavy' subsystem as if it were fixed at a definite location $x_1$, and the `light' subsystem as moving in the resulting potential $\hat W(x_1)$; for instance, the equation could describe how electrons would move for a certain classical configuration of nuclei.\footnote{This step is typically given a heuristic gloss in terms of treating the `heavy' subsystem as if it were `fixed' or `clamped' at a definite location $x_1$, and the `light' subsystem as moving in the resulting potential $\hat W(x_1)$. Physically, however, there is no sense in which the nuclei are literally fixed at points, since they are quantum objects. Rather the step should be understood as a formal move in an approximation scheme, and the `fixed' description a heuristic gloss, which is ultimately both unphysical and unncessary.} As a result (\ref{eq:BO1}) is parameterized by $x_1$, including the eigenvalues (see Figure \ref{fig:BO}) and solutions: hence here $\psi_n(x_1;x_2)$ is a function of $x_2$ (only), of a solution for the potential $\hat W(x_1)$ (hence the semicolon).\\

\BOstep since the $\big(\hat T_2+\hat W(x_1)\big)$ are commuting Hermitian operators for the full system, the $\psi_n(x_1;x_2)$ form an orthonormal basis for states $\Psi(x_1,x_2)$, provided that the observable is non-degenerate. Thus any solution to (\ref{eq:BO1}) can be written
\begin{equation}
\label{eq:BOexact}
    \Psi(x_1,x_2)=\sum_nc_n\theta_n(x_1)\psi_n(x_1;x_2),
\end{equation}
for some (normalized) coefficients $\theta_n(x_1)$. This equation is exact, but now one makes the first approximation and makes a \emph{separation ansatz}: 
\begin{equation}
\label{eq:BO2}
    \Psi(x_1,x_2)\approx\theta_n(x_1)\psi_n(x_1;x_2).
\end{equation}
The rationale for this separation is that because of the separation of masses, there will be far more kinetic energy in the light subsystem than the heavy one, and $E\approx\lambda_n$. In particular, in the regime of stable molecules we have energies and atomic configurations for which the separation between $\lambda_n$s for different values of $n$ are much greater than the kinetic energy of the heavy subsystem: $|\lambda_m-\lambda_n|\gg T_1$. But then (see the Appendix [1]) a superposition of $\psi_n$ cannot be an eigenvector of total energy, and \eqref{eq:BO2} holds.\\

\BOstep it follows from \eqref{eq:BO2} (see the Appendix [2]) that solutions to (\ref{eq:BO1}) have small variation with respect to $x_1$: for molecules, that the electron wavefunction is unchanged by small changes of the nuclei position parameters.

\begin{equation}
    \label{eq:BOapprox}
    \frac{\partial\psi_n(x_1;x_2)}{\partial x_1}\approx0.
\end{equation}
Using this \emph{adiabatic approximation}, an equation for $\theta_n(x_1)$ can be found from (\ref{eq:SETot}), (\ref{eq:BO1}), and (\ref{eq:BO2}):
\begin{equation}
\label{eq:BO3}
    \big(\hat T_1 - (E-\lambda_n))\big)\theta_n(x_1)=0.
\end{equation}
\bigskip

In sum, for molecules, the overall BO method is to solve (\ref{eq:BO1}) and (\ref{eq:BO3}), and then insert into (\ref{eq:BO2}) to yield an approximate solution to \eqref{eq:SETot}. This is a formal technique to construct approximate solutions to a partial differential equation within a well controlled regime of validity. The approximation requires neither time dependent dynamics nor `fixed' nuclear positions for its application.\footnote{The second point is significant for the interpretation of Born-Oppenheimer in the context of the reduction-emergence debate and will be considered in detail in future work. See \cite{woolley:1977,woolley:1978,claverie:1980,hendry:1998,hendry:2006,hendry:2010,hendry:2010b,hendry:2017,fortin:2021,accorinti:2022,chang:2015,gonzalez:2019,cartwright:2022,scerri:2012,hettema:2017,franklin:2020,seifert:2020,seifert:2022}.}  
\bigskip

For the MSS Wheeler-DeWitt equation the rationale -- hence method -- is somewhat different. In particular, we are not interested in finding solutions to \eqref{eq:SETot} for \emph{different} values of $E$, but only for $E=0$. As a result, we are not interested solving \eqref{eq:BO1} for values of $\lambda_n$ -- the energy of the light subsystem --- \emph{widely separated} with respect to the kinetic energy of the heavy subsystem, $T_1$, since these would correspond to different energy levels for the total system. We are concerned with an energy spectrum of a different character than that depicted in Figure \ref{fig:BO}. Hence, (1) we \emph{cannot} justify the separation ansatz \eqref{eq:BO2} of Step 2, since it depends on such energy gaps, and we instead look for superpositions of the form \eqref{eq:BOexact}. This turns out to be important, since finding solutions that plausibly describe classical spacetimes requires forming states from such superpositions, as we shall soon see. Then, (2) we cannot justify the adiabatic approximation of Step (3) \eqref{eq:BOapprox} in terms of the separation ansatz, and a different justification must be given to obtain it and hence \eqref{eq:BO3}.

That is to say that, for quantum cosmology, the BO method is \emph{first} to follow Step 1, and find (normalized) solutions to \eqref{eq:BO1} for the light -- field -- subsystem.
\begin{equation}
\label{reducedWdW}
\big(-\frac{\partial^2}{\partial \phi ^{2}} -ke^{4\alpha} +m^2e^{6\alpha}\phi^2\big) \psi_n (\alpha; \phi) = \lambda_n(\alpha) \psi_n(\alpha;\phi),
\end{equation}
where $x_1$ and $x_2$ are replaced by $\alpha$ and $\phi$, respectively. As before, both the `potential' $\hat W(\alpha) = -ke^{4\alpha} +m^2e^{6\alpha}\phi^2$, and the reduced energy eigenfunctions $\lambda(\alpha)$ are parameterized by the heavy coordinate $\alpha$, which is considered to be a `fixed' c-number parameter rather than a wavefunction variable: \emph{as if} the field were propagating in a fixed spatial geometry $\alpha$. The (exact) solutions are given in \citet[(5.15)]{Kiefer:1988}. As before, they form a basis \eqref{eq:BOexact} for the joint system, and we can write any state as a superposition:
\begin{equation}
    \label{eq:BOexpandQG}
    \Psi(\alpha,\phi)=\sum_nc_n\theta_n(\alpha)\psi_n(\alpha;\phi).
\end{equation}

Then \emph{second} one follows Step 3, and makes the adiabatic approximation 
\begin{equation}\label{eq:BOapprox2}
    \frac{\partial\psi_n(\alpha;\phi)}{\partial \alpha}\approx0,
\end{equation}
and use the Wheeler-DeWitt equation (\ref{WDWmini}) and (\ref{eq:BOexpandQG}) to derive an equation for the wavefunction of the `heavy' degree of freedom:

\begin{equation}\label{eq:SCT}
\big{[} \frac{\partial^2}{\partial^2\alpha} + \lambda_n(\alpha) \big{]}\theta_n(\alpha)=0. 
\end{equation}
Of course, compared to \eqref{eq:BO3} we have $E=0$. Moreover, one can further show that products $\theta_n\psi_n$ are solutions of the Wheeler-DeWitt equation \eqref{WDWmini}, as of course are their superpositions, by linearity.

This equation one solves under a semi-classical Wentzel-Kramers-Brillouin (WKB) approximation with wave-packets peaked about classical trajectories for $\alpha$; we are ultimately interested in understanding the emergence of classical spacetime. We will not investigate here the application of the WKB approximation in the same detail as the Born-Oppenheimer method, but rather refer the reader to \cite[Eq. 5.23]{Kiefer:1988} for the explicit expression for $\theta_n(\alpha)$. It is important for our later analysis that in general terms, the validity of WKB approximation can be understood in terms of a restriction on the \textit{time-independent} characteristic functional of the Hamilton-Jacobi formalism. This, in turn, can be shown to imply that the approximation is valid for regimes in which the de Broglie wavelength is small compared to the characteristic distance over which the spatial potential varies \citep[pp. 112-6]{sakurai:1995}.  (Or, for the Wheeler-DeWitt equation, for regimes in which the wavelength of $\Psi(\alpha,\phi)$ with respect to $\alpha$  is small compared to the distance over which $W(\alpha$) varies.)

What of the required justification of the BO approximation \eqref{eq:BOapprox2}? (Recall, in the molecular case it was justified by the separation ansatz, itself justified by the large electronic energy gaps, which does not now hold.) In the \emph{third} step, once one has solved for $\psi_n(\alpha;\phi)$ and $\theta_n(\alpha)$, then one can simply check by explicit calculation whether \eqref{eq:BOapprox2} holds. \citet[(5.24)]{kiefer:1987} identifies the regime of validity as
\begin{equation}
\label{approxregieme}
n \ll m^{3\alpha},
\end{equation} 
where $n$ is the excitation level of the reduced eigenvalue problem \eqref{reducedWdW}. When this condition is satisfied, the adiabatic approximation holds, and the Born-Oppenheimer method is formally valid. The functions $\theta_n(\alpha)$ and $\psi_n(\alpha;\phi)$ obtained satisfy the equations for the light and heavy parts -- \eqref{eq:BO3} and \eqref{reducedWdW}, respectively -- derived from \eqref{WDWmini} and the adiabatic approximation: so their products are approximate solutions of the Wheeler-DeWitt equation. That is, \eqref{eq:BOapprox2} is an ansatz whose validity can -- \emph{and is} -- checked after putative solutions are found. We will return to this point below.

\subsection{Approximate Derivation of Chronordinal and Chronometric Structure}
\label{approxcocm}

With the general solution found (in the specified regime), the next step is to fix boundary conditions and construct explicit solutions for an emergent FLRW model in which the appropriate semi-classical behaviour is obtained. In particular, we require that we: i) have a well-posed boundary problem for the wavefunction; and ii) it is possible to construct quantum states that form narrow wave-packets that follow the classical solution according the Ehrenfest relation (i.e., with the centroid of the wave-packet approximating the classical behaviour). Significantly, the emergence of such semi-classical behaviour is not guaranteed merely by the validity of the WKB-approximation.    

In Kiefer's model, one finds that the spatially flat, $k=0$, case allows the formulation of a well-posed Cauchy problem for `initial' data surfaces $\alpha_0=\text{constant}$. The full wavefunction \eqref{eq:BOexact} is uniquely determined by a choice of $\Psi(\alpha_0,\phi)$ and $\frac{\partial}{\partial \alpha} \Psi(\alpha,\phi)|_{\alpha=\alpha_0}$. For this case, Kiefer shows that \eqref{eq:SCT} can be solved under the WKB approximation and that, within this approximation, it is possible to construct a wave-packet $\Psi(\alpha,\phi)$ that follows the classical trajectory with minimal dispersion. Specifically, we have a wave packet whose support lies  along a classical FLRW trajectory in MSS; which is uniquely determined by an `initial' wavefunction  $\Psi(\alpha_0,\phi)$; and which is a Gaussian state with respect to $\phi$. Such states are the quantum MSS versions of the quantum coherent states considered to describe approximately classical systems. Then, the solution $\Psi(\alpha_0,\phi)$ to the Wheeler-DeWitt equation \eqref{WDWmini} remains (approximately) Gaussian -- hence classical -- in $\phi$ as $\alpha$ varies, and the peak of the Gaussian follows an FLRW trajectory $\phi(\alpha)$ as per the Ehrenfest relation \cite[pp. 1768-9]{Kiefer:1988}, cf. \cite{kiefer:1990}.  

The spatially flat case allows us to identify the features required for a quantum variable within a Wheeler-DeWitt type model to be identified as an approximate internal time -- that is, to play `classical-time-parameter-like' functional role in the emergent dynamics, as discussed in \S\ref{EmergenceofTime}. The first is for it to be possible to interpret the specification of boundary conditions on the wavefunction as initial (or initial and final) `time' boundary problems. That is, we should be able to fully specify the wavefunction via conditions of the relevant variable and then construct determinate `evolution' of the wavefunction. As we noted, $\alpha$ has this feature. The second feature is that the variable should define a metric for points along this trajectory as, again, $\alpha$ does: $\tau_{ab}\equiv|\alpha_a-\alpha_b|$. Hence, in light of the discussion of \S\ref{EmergenceofTime}, $\alpha$ provides chronometric (hence chronordinal) structure, provided it plays an appropriate temporal role in the emergent dynamics. If Kiefer's model is interpreted as intended, so that $\alpha$ and $\phi$ are the spatial scale factor and a scalar field, respectively, then we say yes: classical FLRW physics is recovered, in which $\alpha$ does play the role of time. 

Note that in this interpretation $\phi$ and $\alpha$ are treated in very different ways: $\phi$ is treated as classical in virtue of being in quantum state corresponding to a classical field. But the wavefunction is \emph{not} localised with respect to $\alpha$ (even for fixed $\phi$), and so is not treated as classical in an approximation to a quantum state. Instead it is \emph{postulated} to correspond to the observed classical time and scale factor; such a correspondence cannot be derived from existing physical principles. That such a new physical postulate will be typically be required for classical spacetime to emerge from quantum gravity is argued in \citet{huggett2021nowhere} and \citet{huggett2021spacetime}; without it the derivation is purely formal, lacking in `physical salience'. Since chronometrical structure does not entail chronodirected structure, this model does not contain an arrow: no basis to say whether the universe is expanding or contracting; for that we need to consider a different model, as we will in \S\ref{sec:approxCD}. 

Whether or not these features obtain in the model depends on the satisfaction of the conditions that allow for the BO and WKB approximations to be applied. In particular, by having a wave packet approximately peaked along the classical trajectory while the condition \eqref{eq:BOapprox2} holds, we know we can label the waveforms in $\{\alpha,\phi\}$ by a parameter that tracks a change that will be, at least locally in configuration space, `continuous' and single valued, namely $\alpha$. By contrast, $\phi$ does not play the functional role of time in the model since the failure of the analogue of condition \eqref{eq:BOapprox2} means that the wave-packet can very rapidly vary along the $\phi$-axis in $\{\alpha,\phi\}$ space. 

It is worth noting that our characterisation of the requirements for a quantum variable to play a functional role a semi-classical time \textit{does not} include the ability to recover an explicit Schr\"odinger-like equation with time dependence with respect to the internal time. One could add such a further role, of course, and this additional role does feature in various semi-classical time approaches to the Wheeler-DeWitt equation, see \cite[p.179]{Kiefer:2012} and references therein. As noted by \cite{chua:2021} this is a plausible \textit{sufficient} condition for a variable to play the functional role of time. However, it is clearly not a \textit{necessary} condition as is shown by the case in hand where such an equation is not derived.

In sum, the overall result is that from the timeless Equation (\ref{WDWmini}), we have extracted an (approximate) \emph{diachronic} structure whereby the internal $\alpha$ degree of freedom plays the functional role of time in the sense of chronometric and chronordinal structures. 

\subsection{Approximate Derivation of Chronodirected Structure }\label{sec:approxCD}

\begin{figure}
    \centering

\centerline{\includegraphics[width=0.8\textwidth]{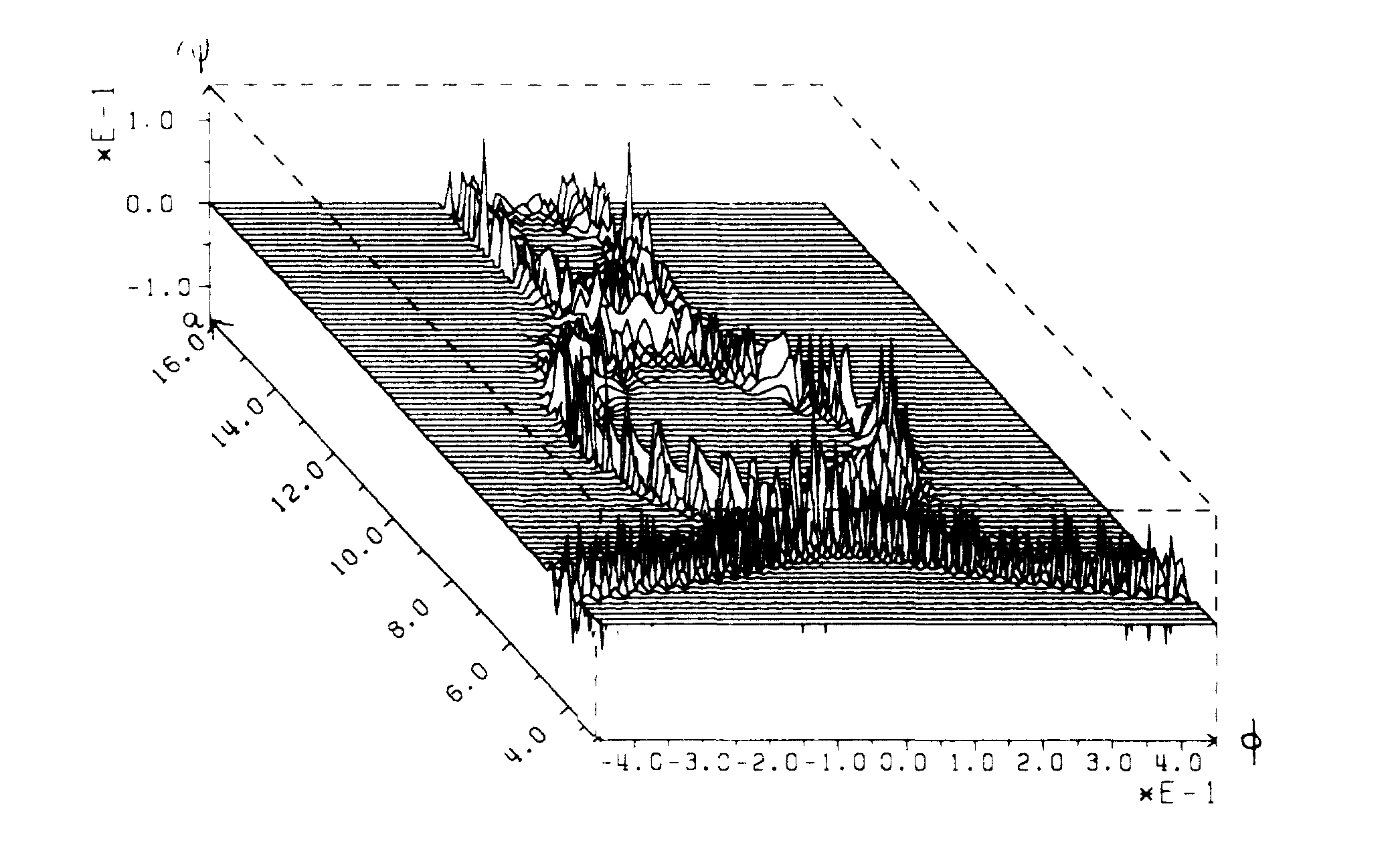}}\centerline{\includegraphics[width=0.8\textwidth]{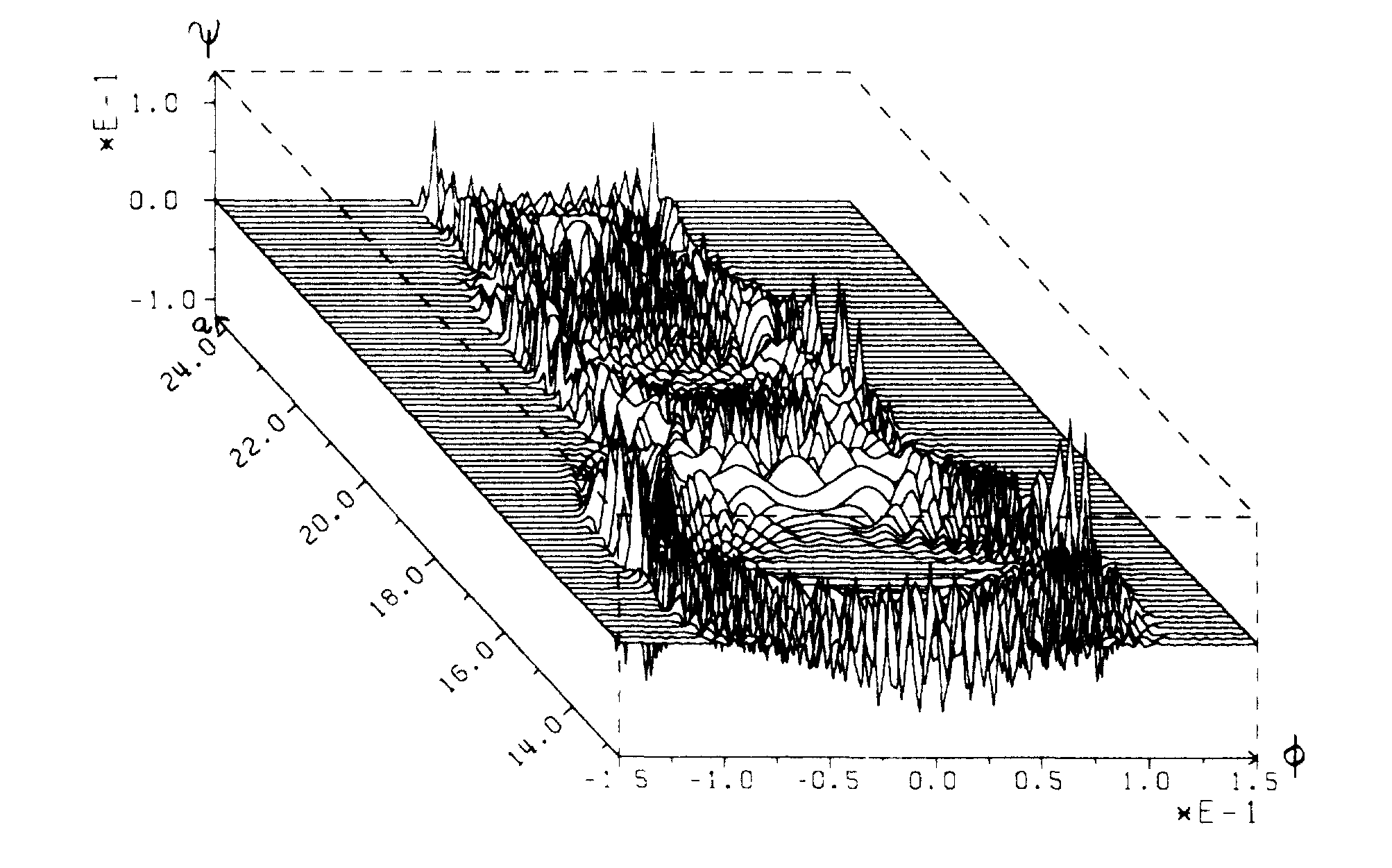}}

    \caption{The wavepacket as a function of the geometric degrees of freedom  ($a=e^\alpha$) and matter degree of freedom ($\phi$) in the case of a closed, $k=+1$, universe. The upper figure shows the region in which a classical trajectory can be recognised with $a$ playing the role of an approximate time parameter. The lower figure shows that at increasing values of $a$, the wave packet begins to spread around the classical trajectory. It is no longer possible to use $a$ as an approximate time parameter and thus the semi-classical emergence of temporal structure fails. (Note the different scales of the plots, which over lap for $14\leq a\leq16$.) \textit{Reprinted figures with permission from Kiefer C. , Physical Review D, 38, 1770, Copyright 1988 by the American Physical Society.}}
    \label{fig:kiefer}

\end{figure}

Of particular relevance to our present discussion is that models of (\ref{WDWmini})  display a \textit{breakdown} of the semi-classical approximation in the case of a closed, $k=+1$, universe. Classical FLRW closed universes with vanishing cosmological constant expand from a big bang then recollapse at a big crunch: the trajectory that we want our Wheeler-DeWitt wavefunction to approximate is a curve in MSS that starts and ends at $\alpha\to-\infty$, and reaches a maximum value of $\alpha$. Therefore, to find a solution to the ($k=+1$) Wheeler-DeWitt equation one therefore: (i) specifies an `initial', $\alpha_0=\text{constant}$, state that is a superposition of expanding \textit{and recollapsing} components; (ii) imposes the boundary condition $\Psi\rightarrow 0$ as $\alpha \rightarrow \infty$; and (iii) seeks dual trajectories approaching a common point as $\alpha$ grows. Such a solution, found by Kiefer, is shown in Figure \ref{fig:kiefer}: particularly the upper plot, of the wavepacket for smaller values of $\alpha$. As before, in such a solution $\alpha$ would provide chronometrical, hence chronordinal structure; though, obviously, that $\alpha$ grows from both the `beginning' and the `end' of the trajectory will affect its identification as time, as we will discuss in \S\ref{asymfromsym}, as we will discuss in \S\ref{asymfromsym}.

The crucial point for now, however, is that (ii) \textit{prevents} the construction of narrow wave-packets with minimal dispersion that would allow us to represent classical trajectories for sufficiently large values of $\alpha$, \cite[p.1770]{Kiefer:1988}, \citep[p.268-9]{Kiefer:2012}. In such circumstances, semi-classical behaviour of the quantum state is not obtained and the putative emergence of temporal structure fails. The failure of putative emergence of temporal structure in the $k=+1$ model is vividly illustrated in the lower plot of Figure \ref{fig:kiefer} which extends the same wavepacket for larger values of $\alpha$.     

Fascinatingly, according to \cite{Kiefer:1988,Kiefer:2012}, the breakdown of the semi-classical limit in this model is precisely the context in which we must invoke the role of the `environment' in inducing classical behaviour. At this point there is an obvious conceptual challenge. On standard approaches to cosmology the universe must be considered a closed system.\footnote{The closed system view is not a physical or logical necessity in cosmology. In particular, open classical statistical systems are standardly defined in terms of conservation of phase space measure and thus the existence of empirically adequate cosmological models based upon contact dynamics which display measure compression indicates that one may model the universe as an open system in the classical domain at least. See \cite{sloan:2021,cuffaro:2021,bravetti:2022,sloan:2023,Ladyman:2023}.} Thus, from this perspective, the role of the environment in a quantum cosmological context must be played by an endogenous system. 

The approach followed in pioneering work by Kiefer and  Zeh is to consider small inhomogeneous degrees of freedom as an endogenous environment and then apply an \textit{internal} version of standard decoherence arguments \citep{zeh:1986,kiefer:1987,Kiefer:1988,zeh:1989,kiefer:1995,Kiefer:2012}.  Recall that the FLRW cosmology is formulated in terms of variables defined on spatial scales in which the universe can be assumed to be homogeneous and isotropic. These `hydrodynamic' variables accurately capture the bulk behaviour of classical spacetimes and are the starting point for building Wheeler-DeWitt type quantum cosmological models. In reality, of course, the universe is not, thankfully, homogeneous and isotropic on all scales. 

A first step towards de-idealisation of a FLRW-type model is to include small inhomogeneities in the classical equations of motion, specifically higher moments of the fields. This is analogous to considering hydrodynamics of bulk fluid variables together with first order fluctuations. This approach can be transferred to the quantum context also. In particular, following \cite[p.347]{Kiefer:2012}, if we write the small inhomogeneous degrees of freedom as $\{x_i\}$ we can arrive at a Wheeler-DeWitt equation of the form:
\begin{equation}
\label{WDWenv}
    \big{[} \frac{\partial^2}{\partial^2\alpha} + \sum_i \frac{\partial^2}{\partial^2 x_i} + V_i(\alpha,x_i) \big{]} \Psi = 0
\end{equation}
where the potentials $V_i(\alpha,x_i)\rightarrow 0$ for $\alpha \rightarrow -\infty$. The form of the potentials means that the Wheeler-DeWitt equation possess an asymmetry with respect to intrinsic time $\alpha$. Furthermore, since the potential is such that the coupling between the bulk variables and inhomogeneities tends towards zero for $\alpha \rightarrow -\infty$ ($a\to0$) one can impose a \textit{Simple Boundary Condition} on the packets of the form (\cite{Conradi:1990rw}):
\begin{equation}
\label{SIC}
    \Psi \xrightarrow[\alpha\rightarrow -\infty]{} \psi_0(\alpha) \prod_i \psi_i(x_i).
\end{equation}
In other words, we assume an `initial state' which (asymptotically) takes the form of a product state in which the bulk variables and inhomogeneities have vanishingly small entanglement.

We then have that if Equation (\ref{SIC}) is taken as an `initial' condition, the Wheeler-DeWitt equation will, through the form of the potential, lead to a wavefunction with an \textit{intrinsic arrow of time}. This would be to derive chronodirected structure in precisely the sense that we described. The crucial formal feature upon which such claims can be justified is that if the intrinsic dynamics of Equation (\ref{WDWenv}) is combined with the simple boundary condition Equation (\ref{SIC}), for increasing $\alpha$ we get increasing entanglement between $\alpha$ and the other modes. 

\cite{Kiefer:2012} (see \S10.3 in particular) interprets this process in terms of an increase in \textit{local entanglement entropy} as defined with reference to a subset of `relevant' degrees of freedom, $\{y_i\}$, where one defines:
\begin{equation}
    S(\alpha, \{y_i\})=-k_B \text{tr} (\rho ln \rho)
\end{equation}
where $\rho$ is the reduced density matrix corresponding to $\alpha$ and $\{y_i\}$.

The final step is then to \textit{interpret} the von Neumann entropy of the reduced density matrix as providing a chronodirected structure. That is, we understand the increase in $S(\alpha, \{y_i\})$ for increasing $\alpha$  as demarcating earlier from later times. As such, it appears that we have a means to derive not just chronordinal and chronometric structure but chronodirected structure also, all starting from a basic equation without an extrinsic time parameter. It is, of course, open to question at this point \textit{why} we interpret the von Neumann entropy as providing chronodirected structure. In essence what we have done is simply stipulate that the arrow of time is equivalent to the von Neumann entropy. Much more is required of the derivation of an arrow of time, including connecting the direction of the von Neumann entropy of the reduced density matrix to the entropy arrows of other physical systems. We will not enter into detailed discussion of this point here, though we will return to related issues in \S\ref{asymfromsym}.  

Let us conclude our discussion by isolating the three crucial assumptions that have gone into the derivation of chronodirected structure in quantum cosmology. The assumptions are:
\begin{itemize}
\item []\textit{Simple Boundary Condition}: The quantum state of the universe, including bulk degrees of freedom and inhomogeneities, at some boundary (initial and final) value of the internal time parameter, can be written \textit{approximately} as a product state, i.e., such that the bulk variables and inhomogeneities have vanishingly small entanglement. 
\item []\textit{Relevant Coarse-Graining}: The degrees of freedom representing inhomogeneities can be divided into relevant and irrelevant subsets and we can represent the state of the universe by a reduced density matrix where the irrelevant subset have been traced out.
\item []\textit{Entropic-Arrow}: Chronodirected structure can be defined via the behaviour of the von Neumann entropy of the reduced density matrix such that an increase in the von Neumann entropy of the reduced density matrix is sufficient to establish a future orientated chronodirected structure.
\end{itemize}
Forms of these assumptions will be familiar from analogous derivations found in statistical mechanics and quantum mechanics. We will return to these connections and the putative justification of the assumptions in \S\ref{asymfromsym}.

\section{Temporal Double Standards}
\label{TDS}

In this section we will consider challenges to various aspects of the approximate derivation of temporal structure detailed in the previous section. The particular focus is a generalisation of the idea of `temporal double standards' -- due to \cite{price:1996} -- in which there is a prima facie worry regarding the cogency of some derivation of temporal structure in physical theory due to implicit appeal to or assumption of such structure within the derivation. We will turn to the specific details of Price's worry relating to the derivation of chronodirected structure in the context in the Kiefer-Zeh approach in \S\ref{asymfromsym}. In \S\ref{notfromt}, we will consider a specific challenge to the  cogency of the work of Kiefer due to \cite{chua:2021}.

\subsection{Time from No Time?}
\label{notfromt}

The reader may have noticed that our presentation of the BO method, for both molecules and quantum cosmology, -- unlike typical pedagogical presentations -- avoided appeal to time dependent physics: for instance, nothing depended on the fact that electrons, being lighter, respond more quickly to perturbations than nuclei. This approach was taken with one eye to responding to a pair of critical challenges due to \cite{chua:2021}.\footnote{It should be noted that  Chua and Callender also provide a very short sketch of a third critical challenge relating to appeal to decoherence. We will here neglect such important considerations, which would inevitably lead into a discussion of the measurement problem, due to restrictions of space.}

The general form of this challenge is concisely stated by the Chua and Callender as follows:
\begin{quote}
Programs in quantum gravity often claim that time emerges from fundamentally timeless physics. In the semiclassical time program, time arises only after approximations are taken. Here we ask what justifies taking these approximations and show that time seems to sneak in when answering this question. This raises the worry that the approach is either unjustified or circular in deriving time from no-time. \citep[p.1172]{chua:2021}
\end{quote}

There are several aspects to this challenge, which we should consider systematically. First, we will have to consider their critique of both the BO and WKB approximations. Second, it is unclear whether the time allegedly sneaking in is external or internal time, since the authors do not explicitly make the distinction in their critique (although they do talk about a `background time metric' as we will see later). If the former then, since there is no external time in canonical quantum gravity, the charge is the devastating one that derivation is incoherent. If the latter, the concern seems weaker: for if one assumes emergent time in one's solutions, and then finds solutions consistent with that assumption, one has shown the existence of emergent time solutions. One has shown that time \emph{can} emerge from no time, but not that it should be expected; something important has been achieved, though one should indeed be clear on its limits. Let us work through these issues one by one.

First then, let us consider the possibility that either the BO or WKB approximations presuppose external time, BO first. Chua and Callender claim that the approximation is `is thoroughly laden with temporal notions' (p.1178) since the light vs. heavy system distinction is a proxy for a slow vs. quick distinction. `The change in the lighter subsystem happens on such a short timescale that there is not enough time for the heavier sub-system to react in that relevant timescale, and so it is effectively independent of lighter subsystems in that period of time' (1178). They assume this to be the case in the context of the molecular version of the approximation and then consider the possibility that it is also the case in the quantum cosmology version. Let us suppose the time scale in question to be extrinsic, so that the putative issue is one of incoherence. 

But our explication shows that in the molecular case the central claim is incorrect (or at least unnecessary): the mass separation is rather proxy for the electronic energy level separation, so a property of solutions of the time-\emph{in}dependent Schr\"odinger equation. Now it is true that textbook presentations do justify the BO approximation along the lines given by Callender and Chua; for instance, the highly influential \citet[XVIII.iii]{messiah:1962} essentially uses it to justify the explicit use of the adiabatic approximation (less explicitly, the move is also made in \citet[\S14]{born1955dynamical}). But this is for heuristic convenience not necessity, as our presentation -- and indeed that of Born and Oppenheimer -- shows.\footnote{To be clear, we do not claim originality for our demonstration, which draws on \citet{Jecko:2014} to reframe \citet{bo:1927}. Note that Messiah discusses the relation between his approach and that of Born and Oppenheimer.} 

For molecules, the Born and Oppenheimer approximation rests upon the relative size of terms in a time-independent Hamiltonian and the consequent splitting of the energy spectrum, which implies separability and adiabaticity. These features justify the approximation, but do not require temporal considerations, though of course they imply dynamical properties too, through the time-\emph{de}pendent Schr\"odinger equation. Things are somewhat different for quantum cosmology, since the splitting of eigenstates does not occur in the desired solutions; so in this case putative solutions obtained by the BO method must be checked formally to verify that they do indeed satisfy the conditions of the approximation -- which they do, in a specified regime, as we saw. One may consider the relation of the time-independent properties to time-scales and the intuitive dynamical picture \textit{if} the Hamiltonian can be taken as the generator of time translations; such an assumption is possible, but \textit{unnecessary} in the molecular context and \textit{inconsistent} in the Wheeler-DeWitt context if the Hamiltonian constraint is interpreted as a `gauge generator' as in the standard interpretation that follows \cite{Henneaux:1992a}.\footnote{For detailed arguments towards the view (not defended here) that the standard treatment is problematic see \cite{Gryb:2024}.} 

These reflections point to the futility of arguing that \emph{any} computation can only be understood by invoking a time-dependent property of a quantum system. For such a property supervenes on the Hamiltonian, and hence on the energy eigenstates and eigenvalues, and hence on the time-independent Schr\"odinger equations -- as these examples illustrate.\footnote{At least for time independent Hamiltonians. A similar argument can be run for classical systems via the Hamilton-Jacobi formalism. In particular, for time-independent classical Hamiltonians, the separation anzatz  for the principal functional in terms of the time-independent characteristic functional together with an $Et$ term establishes an analogous supervenience claim.}

Therefore, because the BO approximation can be justified without appeal to temporal properties, it most certainly does not assume an appeal to an external time, and so its use in quantum cosmology is perfectly coherent. What about the WKB approximation? In this case, Chua and Callender recognise the prima facie atemporality of the approximation. In particular, the fact that the crucial feature is spatial smoothness of the potential function which can be expressed more precisely in terms of de Broglie wavelength being small compared to the characteristic distance over which the spatial potential varies.  However, they contend:

\begin{quote}
Still, time is present. There are many ways to see this. [...] In quantum mechanics, we know that the momentum operator depends only on spatial variables and not time [...] However, the classical momentum does depend on time since $
p=m\frac{dx}{dt}$ [...] The time dependence, evident when talking about velocities/momenta, becomes masked when we replace velocities with notions of wavelengths and spatial variations. Yet the time dependence is plainly there. From [the de Broglie relation and dimensional analysis] if [the scale of variation of the potential] is large, the WKB approximation will be good for long [time scales] and if small then only for short  [time scales]. In standard cases WKB is thus justified via a background time metric. In the case of semiclassical time, however, there is no such background time metric, so we again face our challenge to justify the assumption without invoking time. \cite[pp. 1179-81]{chua:2021}
\end{quote}

What should we make of these claims? Most straightforwardly, it is in general incorrect that the classical canonical momenta variables are given by $m\frac{dx}{dt}$. This is an expression for a particular family of Newtonian mechanical theories in a particular choice of canonical coordinates. In the general context, $p$ is a generalised canonical momentum variable given by an element of the phase space. In particular, we have that $p\in\Gamma$ where $(\Gamma,\omega)$ is the symplectic manifold given by the cotangent bundle to configuration space $\Gamma=T^\star\mathcal{C}$. Canonical momenta are only defined up to to symplectomorphism, and privileging a representation in a preferred chart is inconsistent with mathematical practice regarding their representational roles, cf. \cite{weatherall:2018,gryb:2016}. Moreover, even in an explicit coordinate representation, the momenta are in fact given by $p=\frac{\partial L}{\partial \dot{x}}$ and we only recover the Newtonian momentum given a particular form of the Lagrangian. The required temporal structure is that of infinitesimal tangent and cotangent vectors, not a background temporal metric. 

Furthermore, as with the case of Born-Oppenheimer, the intuitive justificatory story told for the quantum mechanics example that does feature temporal notions, is  \textit{unnecessary}. The WKB approximation in quantum theory can be fully justified in a time independent context since its validity is a property of a spatial function. This, then, also holds for the case WKB in the context of Wheeler-DeWitt quantum cosmology. Here we can think of the formal requirements in terms of in terms needing some notion of functional derivation in the space of Riemannian three metrics which is not itself a temporal structure.\footnote{Thanks to Henrique Gomes for pointing this out to us.}  

Thus, we conclude, there is nothing to the potentially damning complaint that the derivation of time from no time assumes the existence of an external time. However, there remains the charge that the derivation is circular in the sense that it assumes an internal time. In response, we observe that the formal structure of the Kiefer's derivation is that of an \emph{ansatz}, the standard mathematical approach where a trial form of solution to a differential equation is assumed and then tested for consistency. Separability and adiabaticity are assumed, and the solutions checked to see that they obey the condition. Indeed, Born and Oppenheimer's original derivation used the very same logic. Specifically, they had to assume that in the lowest order solutions the nuclei were at the minima of their potentials: `\textit{eine Annahme, die erst durch den Erfolg gerechtfertigt werden kann}' (`an assumption only justified by its success') \cite[p.465]{bo:1927} (trans. Blinder). Similarly, in Kiefer's derivation we in fact find an explicit estimation of the regime of validity of the BO approximation in \eqref{approxregieme}. Moreover, we have seen that the WKB approximation holds in some regimes, and not in others. There is no uncontrolled circularity here.

The problem at this point is that there is no formal or physical restriction that prohibits the existence of solutions to our Wheeler-DeWitt equation \eqref{WDWmini}, which \textit{do not} satisfy the conditions, and in which time is thus not emergent. We need to provide a formal or physical justification for the expectation that the approximation conditions are satisfied -- and also the simple boundary condition, for chronodirected structure -- otherwise what has been achieved is a merely a demonstration of possibility, not necessity: that some, but not all, solutions admit emergent time.\footnote{A more general and more worrying possibility comes from the fact that in totally constrained systems, like general relativity, non-integrable dynamics may lead to a quantization that does not admit a semi-classical limit \citep{dittrich:2017}. The story provided here regarding the semi-classical emergence of time would be inadmissible in such circumstances.}

A more ambitious temporal emergentist target would be the emergence of the full classical temporal structure of general relativity from the Wheeler-DeWitt equation. That is, the application of a WKB-type derivation as a bridge for the approximate derivation of the Einstein Field Equations from the Wheeler-DeWitt equation, via the relevant generally relativistic Hamilton-Jacobi equation due to \cite{Peres:1962}. We leave analysis of this full story of the putative emergence of time to future work. Discussion of the relevant derivations in the physics literature can be found in \cite{Gerlach:1969,Komar:1971,Misner:1974,hartle:1993,kiefer:2009,Salisbury:2020}.

\subsection{Asymmetry from Symmetry?}
\label{asymfromsym}

In the previous subsection we considered potential issues for the emergence of chronordinal and chronometric structure in the context of their approximate derivation from a Wheeler-DeWitt cosmology, so without such structures at a basic level. The challenges accuse the derivation of circularity in some form -- somehow assuming the very temporal structure to be derived. In the context of the derivation of chronodirected structure (with chronordinal structure already given) an analogous challenge was made in the now classic discussion of \cite[chapters 2-4]{price:1996}, in which a `temporal double standards' argument pattern is repeated across multiple physical contexts from thermal physics to quantum cosmology. We can rationally reconstruct it as follows:

\begin{enumerate}
    \item Suppose theory $T$ possesses chronordinal structure, with respect to which the dynamics is time symmetric: the time reverse of any solution is another solution.
    \item So the time-asymmetry of the world -- its chronodirected structure -- could only be explained from $T$ via appeal to a special boundary condition.
    \item But to apply such a boundary condition to \emph{only} one end of time is to indulge in a temporal double standard (which may not be obvious until we take the appropriate `view from nowhen').\footnote{Of course, supposing that the boundary condition can consistently hold at both ends of time, as indeed it can in all the relevant cases; for instance, entropy could be low at either end of a chronordinal structure.}
    \item Then, the chronodirectedness of the world has either not been explained at all or has been put in by hand based upon a temporal double standard.
\end{enumerate}
In this section we will consider how the derivation of chronodirected structure in quantum cosmology deals with this powerful and important challenge.

To start with, observe a near intersection between our case study of Kiefer and Zeh in \S\ref{sec:approxCD} with Price's own discussion of the putative emergence of time in the famous `no-boundary proposal' of \cite{hartle:1983}. Following the argument pattern reconstructed above, Price takes particular issue with Hawking and coauthors' proposal that the no-boundary condition be applied to only one `end' of the universe, so that chronodirected structure emerges only in one temporal orientation \citep{hawking:1993,hawking:1994}. (This is in contrast to Hawking's original position expressed in \citep{Hawking:1985}, in which the arrow of time would `reverse' in a re-collapsing universe and the no-boundary proposal would apply at the limit of both temporal extremities.)

\begin{quote}
    [Hawking] hasn't shown that asymmetric universes are the natural product of a symmetric theory [...] on the contrary,  he seems to have simply assumed the required asymmetry, by taking the no boundary condition to apply to only one end of an arbitrary universe [...] this amounts to putting the asymmetry in ``by hand.'' \cite[p.93]{price:1996}
\end{quote}
We concur.

To what extent, then, does the emergence of chronodirected structure based on the simple boundary condition of Kiefer and Zeh also amount to `putting the asymmetry in by hand'? Answering this question carefully will help us obtain a clearer picture of the arrow of time in the semi-classical closed universe model. 

First, with respect to (1) of Price's double standards argument, the laws are not time symmetric; indeed, they are neither (fundamentally) temporal nor symmetric! The Wheeler-DeWitt equation is atemporal, but with an asymmetry with respect to $\alpha$ in the form of the potential. Second, therefore the simple boundary condition (\ref{SIC}) is not postulated as an \textit{initial} condition, but a \textit{boundary} condition that applies for $\alpha \rightarrow -\infty$: that the initial state is separable, in the regime in which the matter and gravitational subsystems are non-interacting. Third, in the semi-classical closed universe model $\alpha \rightarrow -\infty$ at both `ends', so in the sense of emergent time, the condition (\ref{SIC}) is applied symmetrically, as Price insists against Hawking et al. Of course, given the identification of a future-past arrow with the direction of increasing (von Neumann) entropy, this double imposition of the boundary condition, entails that the arrow of time points towards increasing $\alpha$ in \emph{both} branches coming from $\alpha \rightarrow -\infty$.

Although he does not discuss Kiefer and Zeh's work\footnote{Their paper contains many of the same criticisms of Hawking.}, \citet[pp.99-111]{price:1996} defends the possibility of a `Gold universe' -- i.e., one with low entropy at the start and end of time, and hence two, opposite arrows of time, one before and one after a `turn-around', just as in our case. He further argues that in such a universe one could now observe traces of events in the post-turn-around future: after all, with respect to the oppositely oriented temporal arrow at such an event, the event is in our past! However, Price's reasoning assumes a classical spacetime background, and so does not apply to the Kiefer-Zeh model, since at the turn-around, where the two branches meet, the system becomes fully quantum \cite[p.4150]{kiefer:1995} and, as Wheeler said earlier, `there is no spacetime'.

That is, the model provides a mechanism for the chronodirected structure to \textit{disappear} as well as emerge. Since there is no classical connection between different `legs' across the turning point between expansion and collapse, rather we find, at large $a$, a region which cannot be interpreted in classical terms at all. Relative to an observer located in a semi-classical portion of the universe, the full story of temporal structure is then one of emergence, disappearance, re-emergence, and then at a final big crunch, disappearance again. 

\begin{figure}\label{fig:janus}
    \centering
    \includegraphics[width=4in]{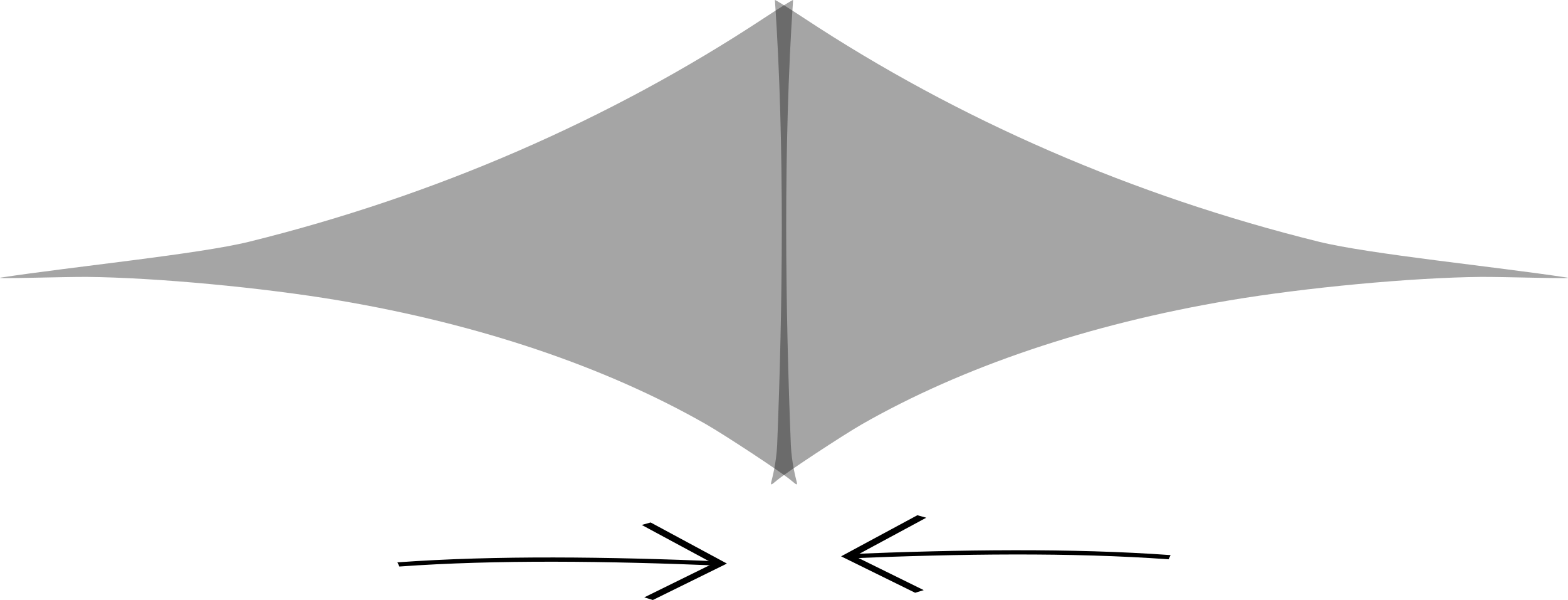}
    \caption{The view from `nowhen' of the double-universe. Where the temporal arrows meet, spacetime breaks down.}
    \label{fig:enter-label}
\end{figure}

However, because the emergent arrow of time reverses in the model, observers on either side of the turn-around give the same description, even though it happens in opposite relative temporal directions! Thus it is more perspicuous to take Price's `view from nowhen' again. Consider: (a) the simple condition sets the entropy at a minimum when the universe is small, and (b) we identify the future as the direction in which entropy increases, which it will (c) as the interaction between bulk degrees of freedom and inhomogeneities grows producing entanglement, (d) which it does as $\alpha$ grows -- i.e., as the universe expands. That is, the simple condition, the identification, the Wheeler-DeWitt dynamics, and the form of the Wheeler-DeWitt potential entail that the universe can only expand, and never contract. We might thus, more consistently atemporally, say that the arrow of time does not \textit{reverse}, and the universe does not \textit{recollapse} in the model. Rather the model describes a `transition' between two semi-classical universes, each with an emergent arrow of time orientated away from the low volume regime. The arrow in question, when it is well defined, \textit{always} points away from the $\alpha \rightarrow -\infty$ regime and towards the $\alpha \rightarrow +\infty$ regime, and it is thus, by definition, aligned with the expansion of the universe. However, their temporal structure dissolves as they merge into a large volume deep quantum regime (see Figure \ref{fig:janus}. Whilst one could hardly wish for a more symmetric treatment, the physical interpretation of the breakdown of classical physics at large scales is surely more than a little counter-intuitive.

Let us finally turn to the question of physically justifying the three assumptions needed for the Kiefer-Zeh derivation of chronordinal structure in quantum cosmology -- and so justify the claim that temporal structure is not merely derived, but emergent, according to the definition we gave. The first was the simple boundary condition together with the relevant form of the potential. It was assumed that the quantum state of the universe, including bulk degrees of freedom and inhomogeneities, at some boundary (initial and final) value of the internal time parameter, can be written approximately as a product state, i.e., such that the bulk variables and inhomogeneities have vanishingly small entanglement. Here it is surely fair to conclude that the story provided by Kiefer and Zeh is ultimately not a satisfactory one. In particular, there is no clear \textit{physical justification} for the simple boundary condition or the asymmetry of the potential in $\alpha$. However, in this regard, the situation is not notably worse than that of the famous `past hypothesis' -- and indeed the special boundary condition (\ref{SIC}) is a form of low entropy initial state that can be related to a quantum version of Penrose's Weyl curvature hypothesis \citep{kiefer:2022b}. 

In this context, one might, then simply reject the demands for further justification, or indeed \textit{explanation}, on the grounds that we do not typically expect initial conditions to be explained \citep{callender2004measures}. However, following \cite{price2004origins} once more, one might question whether universal boundary conditions are relevantly similar to local boundary conditions; if not, Callender's analogy fails. Then again, if `initial' conditions like the Kiefer-Zeh simple boundary condition should be thought of as ‘law-like’ rather than ‘fact-like’, once again the demand for explanation arguably vanishes. There are clearly philosophical issues to be addressed, which are outside the scope of this work, but the obvious conclusion is that the physical origin of such conditions is an important and incomplete explanatory project, which we would do well to pursue further.

The next crucial assumption was \textit{relevant} coarse-graining. It was assumed that the degrees of freedom representing inhomogeneities can be divided into relevant and irrelevant subsets and that we can represent the state of the universe by a reduced density matrix where the irrelevant subset has been traced out. In this context, it is possible to draw upon parallel discussions in the context of statistical mechanics. In particular, consider the insightful analysis of \cite{robertson:2020}\footnote{See \cite{Wallace:2011} and \cite{zeh:1989} for related ideas. For work on coarse-graining in classical cosmology see \cite{te:2021}. For a quantum cosmological model studied in the inflationary context see \cite{hollowood:2017}.}: according to Robertson, the crucial feature of an \textit{objective} coarse-graining, i.e. one that is not illusory or problematic anthropocentric, is that it allows us to abstract to a higher-level \textit{autonomous} description. Autonomy here is understood in the dynamical systems sense of having independently well-posed evolution equations. 

Similarly, one could argue that the coarse-grained asymmetry derived by Kiefer and Zeh provides an objective basis for the emergence of chronodirected structure. To do so would involve showing that the irrelevant inhomogeneous degrees of freedom are such that the dynamics of the rest of the universe is suitably autonomous. Such a project is attempted, with some but not complete success, in \citet{kiefer:1987}, following \citet{halliwell:1985}.

The third and final assumption is the entropic-arrow assumption which amounts to the idea that chronodirected structure can be defined the von Neumann entropy of the reduced density matrix. In this context, one can note that the monotonic (or secular) increase of a entropy function does not in fact require or enforce chronodirected structure but can instead be perfectly compatible with a theory of time in which only chronordinal structure obtains \citep{farr:2016,farr2020c}. Moreover, if the aim in providing approximate derivation of chronodirected structure was to provide an explanation for the full gamut of temporally directed phenomena required by our functionalism, it is not clear that models such as that of Kiefer and Zeh are capable of addressing such an  ambitious explanatory project, cf. \cite{Ryder:2022}. In particular, no resources have been provided to connect the derived global entropic arrow with local temporally directed processes, such as equilibration of thermal systems or memory formation. In this sense, the model, although remarkable, still falls short of explaining the arrow of time, until and unless such connections are provided.

\section{Conclusion}

In this paper we have conducted a case study analysis of a proposal for the emergence of time based upon the approximate derivation of three grades of temporal structure within an explicit quantum cosmological model which obeys a Wheeler-DeWitt type equation without an extrinsic time parameter. Our conclusion is that the model provides a self-consistent account of the emergence of chronordinal, chronometric and chronodirected structure.

Our analysis has demonstrated that the Born-Oppenheimer and WKB approximations can be justified without presupposing  the chronordinal and chronometric structures that one is aiming to derive. Further, regarding emergent chronodirected structure, we have shown that the Kiefer-Zeh model avoids Price's temporal double standards argument due to its temporal symmetry. In particular, since the model includes two symmetrically situated emergent arrows of time, it allows us to avoid asymmetry by not having an end the universe, but rather two beginnings. There is evidently an exciting explanatory project with regard to the emergence of chronodirected structure that should be actively pursued in the context of more realistic cosmological models. Moreover, the even more difficult challenge remains of interpreting the model in the context of the cosmological measurement problem.

\section*{Acknowledgements}

We are extremely appreciative to audiences in Buenos Aires, Surrey, Ovronnaz, and Milan for feedback, and to  Henrqiue Gomes, James Ladyman, Huw Price and especially Claus Kiefer for comments and discussion. Work on this project was supported by a Bristol Benjamin Meaker Distinguished Visiting Professorship, grant IZSEZ0\_214119 from the Swiss NSF, a Visiting Fellowship at Magdalen College, Oxford, and by John Templeton Foundation grant numbers 61387 and 62210. The views expressed may not represent the opinions of the Foundation.

\bibliographystyle{chicago}
\bibliography{Masterbib3.bib,singularbib.bib,SD,philbib}

\begin{thebibliography}{}

\bibitem[\protect\citeauthoryear{Accorinti and Gonz{\'a}lez}{Accorinti and
  Gonz{\'a}lez}{2022}]{accorinti:2022}
Accorinti, H.~L. and J.~C.~M. Gonz{\'a}lez (2022).
\newblock Models and idealizations in quantum chemistry: The case of the
  born-oppenheimer approximation.
\newblock {\em Philosophical Perspectives in Quantum Chemistry\/}, 107--124.

\bibitem[\protect\citeauthoryear{Arnowitt, Deser, and Misner}{Arnowitt
  et~al.}{1960}]{ADM:1960}
Arnowitt, R., S.~Deser, and C.~W. Misner (1960, Mar).
\newblock Canonical variables for general relativity.
\newblock {\em Phys. Rev.\/}~{\em 117}, 1595--1602.

\bibitem[\protect\citeauthoryear{Ashtekar and Bianchi}{Ashtekar and
  Bianchi}{2021}]{ashtekar:2021}
Ashtekar, A. and E.~Bianchi (2021).
\newblock A short review of loop quantum gravity.
\newblock {\em Reports on Progress in Physics\/}.

\bibitem[\protect\citeauthoryear{Ashtekar, Pawlowski, and Singh}{Ashtekar
  et~al.}{2006}]{ashtekar:2006a}
Ashtekar, A., T.~Pawlowski, and P.~Singh (2006).
\newblock Quantum nature of the big bang.
\newblock {\em Physical review letters\/}~{\em 96\/}(14), 141301.

\bibitem[\protect\citeauthoryear{Ashtekar and Singh}{Ashtekar and
  Singh}{2011}]{ashtekar:2011}
Ashtekar, A. and P.~Singh (2011).
\newblock Loop quantum cosmology: a status report.
\newblock {\em Classical and Quantum Gravity\/}~{\em 28\/}(21), 213001.

\bibitem[\protect\citeauthoryear{Banks}{Banks}{1985}]{banks:1985}
Banks, T. (1985).
\newblock Tcp, quantum gravity, the cosmological constant and all that...
\newblock {\em Nuclear Physics B\/}~{\em 249\/}(2), 332--360.

\bibitem[\protect\citeauthoryear{Barbero~G and Villase{\~n}or}{Barbero~G and
  Villase{\~n}or}{2010}]{barbero:2010}
Barbero~G, J.~F. and E.~J. Villase{\~n}or (2010).
\newblock Quantization of midisuperspace models.
\newblock {\em Living Reviews in Relativity\/}~{\em 13\/}(1), 1--55.

\bibitem[\protect\citeauthoryear{Black}{Black}{1959}]{black:1959}
Black, M. (1959).
\newblock The" direction" of time.
\newblock {\em Analysis\/}~{\em 19\/}(3), 54--63.

\bibitem[\protect\citeauthoryear{Blyth and Isham}{Blyth and
  Isham}{1975}]{blyth:1975}
Blyth, W. and C.~Isham (1975).
\newblock Quantization of a friedmann universe filled with a scalar field.
\newblock {\em Physical Review D\/}~{\em 11\/}(4), 768.

\bibitem[\protect\citeauthoryear{Bojowald}{Bojowald}{2001}]{bojowald:2001}
Bojowald, M. (2001).
\newblock Absence of a singularity in loop quantum cosmology.
\newblock {\em Physical Review Letters\/}~{\em 86\/}(23), 5227.

\bibitem[\protect\citeauthoryear{Born, Huang, and Lax}{Born
  et~al.}{1955}]{born1955dynamical}
Born, M., K.~Huang, and M.~Lax (1955).
\newblock Dynamical theory of crystal lattices.
\newblock {\em American Journal of Physics\/}~{\em 23\/}(7), 474--474.

\bibitem[\protect\citeauthoryear{Born and Oppenheimer}{Born and
  Oppenheimer}{1927}]{bo:1927}
Born, M. and R.~Oppenheimer (1927).
\newblock Zur quantentheorie der molekeln.
\newblock {\em Annalen der Physik\/}~{\em 389\/}(20), 457--484.

\bibitem[\protect\citeauthoryear{Bravetti, Jackman, and Sloan}{Bravetti
  et~al.}{2022}]{bravetti:2022}
Bravetti, A., C.~Jackman, and D.~Sloan (2022).
\newblock Scaling symmetries, contact reduction and poincar$\backslash$'e's
  dream.
\newblock {\em arXiv preprint arXiv:2206.09911\/}.

\bibitem[\protect\citeauthoryear{Butterfield}{Butterfield}{2011}]{butterfield:2011}
Butterfield, J. (2011).
\newblock Less is different: Emergence and reduction reconciled.
\newblock {\em Foundations of physics\/}~{\em 41}, 1065--1135.

\bibitem[\protect\citeauthoryear{Callender}{Callender}{2004}]{callender2004measures}
Callender, C. (2004).
\newblock Measures, explanations and the past: Should `special' initial
  conditions be explained?
\newblock {\em The British journal for the philosophy of science\/}~{\em
  55\/}(2), 195--217.

\bibitem[\protect\citeauthoryear{Callender}{Callender}{2021}]{sep-time-thermo}
Callender, C. (2021).
\newblock {Thermodynamic Asymmetry in Time}.
\newblock In E.~N. Zalta (Ed.), {\em The {Stanford} Encyclopedia of
  Philosophy\/} ({S}ummer 2021 ed.). Metaphysics Research Lab, Stanford
  University.

\bibitem[\protect\citeauthoryear{Cartwright}{Cartwright}{2022}]{cartwright:2022}
Cartwright, N. (2022).
\newblock {\em A Philosopher Looks at Science}.
\newblock Cambridge University Press.

\bibitem[\protect\citeauthoryear{Chang}{Chang}{2015}]{chang:2015}
Chang, H. (2015).
\newblock Reductionism and the relation between chemistry and physics.
\newblock {\em Relocating the history of science: Essays in honor of Kostas
  Gavroglu\/}, 193--209.

\bibitem[\protect\citeauthoryear{Chua and Callender}{Chua and
  Callender}{2021}]{chua:2021}
Chua, E.~Y. and C.~Callender (2021).
\newblock No time for time from no-time.
\newblock {\em Philosophy of Science\/}~{\em 88\/}(5), 1172--1184.

\bibitem[\protect\citeauthoryear{Claverie and Diner}{Claverie and
  Diner}{1980}]{claverie:1980}
Claverie, P. and S.~Diner (1980).
\newblock The concept of molecular structure in quantum theory: interpretation
  problems.
\newblock {\em Israel Journal of Chemistry\/}~{\em 19\/}(1-4), 54--81.

\bibitem[\protect\citeauthoryear{Conradi and Zeh}{Conradi and
  Zeh}{1991}]{Conradi:1990rw}
Conradi, H.~D. and H.~D. Zeh (1991).
\newblock {Quantum cosmology as an initial value problem}.
\newblock {\em Phys. Lett. A\/}~{\em 154}, 321--326.

\bibitem[\protect\citeauthoryear{Cuffaro and Hartmann}{Cuffaro and
  Hartmann}{2021}]{cuffaro:2021}
Cuffaro, M.~E. and S.~Hartmann (2021).
\newblock The open systems view.
\newblock {\em arXiv preprint arXiv:2112.11095\/}.

\bibitem[\protect\citeauthoryear{de~Blas, Olmedo, and Paw{\l}owski}{de~Blas
  et~al.}{2017}]{deBlas:2017}
de~Blas, D.~M., J.~Olmedo, and T.~Paw{\l}owski (2017).
\newblock Loop quantization of the gowdy model with local rotational symmetry.
\newblock {\em Physical Review D\/}~{\em 96\/}(10), 106016.

\bibitem[\protect\citeauthoryear{DeWitt}{DeWitt}{1967}]{DeWitt:1967}
DeWitt, B. (1967).
\newblock Quantum theory of gravity. i. the canonical theory.
\newblock {\em Physical Review\/}~{\em 160}, 1113--1148.

\bibitem[\protect\citeauthoryear{Dirac}{Dirac}{1958}]{Dirac:1958b}
Dirac, P. A.~M. (1958).
\newblock The theory of gravitation in hamiltonian form.
\newblock {\em Proceedings of the Royal Society of London. Series A,
  Mathematical and Physical Sciences\/}~{\em 246}, 333--343.

\bibitem[\protect\citeauthoryear{Dirac}{Dirac}{1964}]{Dirac:1964}
Dirac, P. A.~M. (1964).
\newblock {\em Lectures on quantum mechanics}.
\newblock Dover Publications.

\bibitem[\protect\citeauthoryear{Dittrich, H{\"o}hn, Koslowski, and
  Nelson}{Dittrich et~al.}{2017}]{dittrich:2017}
Dittrich, B., P.~A. H{\"o}hn, T.~A. Koslowski, and M.~I. Nelson (2017).
\newblock Can chaos be observed in quantum gravity?
\newblock {\em Physics Letters B\/}~{\em 769}, 554--560.

\bibitem[\protect\citeauthoryear{Farr}{Farr}{2012}]{Farr:2012}
Farr, M. (2012).
\newblock {\em Towards a C Theory of Time}.
\newblock Ph.\ D. thesis, University of Bristol.

\bibitem[\protect\citeauthoryear{Farr}{Farr}{2016}]{farr:2016}
Farr, M. (2016).
\newblock Causation and time reversal.
\newblock {\em The British Journal for the Philosophy of Science\/}.

\bibitem[\protect\citeauthoryear{Farr}{Farr}{2020}]{farr2020c}
Farr, M. (2020).
\newblock C-theories of time: On the adirectionality of time.
\newblock {\em Philosophy Compass\/}~{\em 15\/}(12), e12714.

\bibitem[\protect\citeauthoryear{Fortin and Lombardi}{Fortin and
  Lombardi}{2021}]{fortin:2021}
Fortin, S. and O.~Lombardi (2021).
\newblock Is the problem of molecular structure just the quantum measurement
  problem?
\newblock {\em Foundations of Chemistry\/}~{\em 23\/}(3), 379--395.

\bibitem[\protect\citeauthoryear{Franklin and Seifert}{Franklin and
  Seifert}{2020}]{franklin:2020}
Franklin, A. and V.~A. Seifert (2020).
\newblock The problem of molecular structure just is the measurement problem.

\bibitem[\protect\citeauthoryear{Gerlach}{Gerlach}{1969}]{Gerlach:1969}
Gerlach, U.~H. (1969, Jan).
\newblock Derivation of the ten einstein field equations from the semiclassical
  approximation to quantum geometrodynamics.
\newblock {\em Phys. Rev.\/}~{\em 177\/}(5), 1929--1941.

\bibitem[\protect\citeauthoryear{Gielen and Men{\'e}ndez-Pidal}{Gielen and
  Men{\'e}ndez-Pidal}{2022}]{gielen:2022}
Gielen, S. and L.~Men{\'e}ndez-Pidal (2022).
\newblock Unitarity, clock dependence and quantum recollapse in quantum
  cosmology.
\newblock {\em Classical and Quantum Gravity\/}.

\bibitem[\protect\citeauthoryear{Gonz{\'a}lez, Fortin, and
  Lombardi}{Gonz{\'a}lez et~al.}{2019}]{gonzalez:2019}
Gonz{\'a}lez, J. C.~M., S.~Fortin, and O.~Lombardi (2019).
\newblock Why molecular structure cannot be strictly reduced to quantum
  mechanics.
\newblock {\em Foundations of Chemistry\/}~{\em 21}, 31--45.

\bibitem[\protect\citeauthoryear{Gotay and Isenberg}{Gotay and
  Isenberg}{1980}]{gotay:1980}
Gotay, M.~J. and J.~A. Isenberg (1980).
\newblock Geometric quantization and gravitational collapse.
\newblock {\em Physical Review D\/}~{\em 22\/}(2), 235.

\bibitem[\protect\citeauthoryear{Gryb and Th{\'e}bault}{Gryb and
  Th{\'e}bault}{2016}]{gryb:2016}
Gryb, S. and K.~P. Th{\'e}bault (2016).
\newblock Regarding the `hole argument'and the `problem of time'.
\newblock {\em Philosophy of Science\/}~{\em 83\/}(4), 563--584.

\bibitem[\protect\citeauthoryear{Gryb and Th{\'e}bault}{Gryb and
  Th{\'e}bault}{2018}]{gryb:2018}
Gryb, S. and K.~P. Th{\'e}bault (2018).
\newblock Superpositions of the cosmological constant allow for singularity
  resolution and unitary evolution in quantum cosmology.
\newblock {\em Physics Letters B\/}~{\em 784}, 324--329.

\bibitem[\protect\citeauthoryear{Gryb and Th{\'e}bault}{Gryb and
  Th{\'e}bault}{2011}]{gryb:2011}
Gryb, S. and K.~P.~Y. Th{\'e}bault (2011).
\newblock The role of time in relational quantum theories.
\newblock {\em Foundations of Physics\/}, 1--29.

\bibitem[\protect\citeauthoryear{Gryb and Th{\'e}bault}{Gryb and
  Th{\'e}bault}{2014}]{gryb:2014}
Gryb, S. and K.~P.~Y. Th{\'e}bault (2014).
\newblock Symmetry and evolution in quantum gravity.
\newblock {\em Foundations of Physics\/}~{\em 44\/}(3), 305--348.

\bibitem[\protect\citeauthoryear{Gryb and Th\'{e}bault}{Gryb and
  Th\'{e}bault}{2016a}]{Gryb:2016a}
Gryb, S. and K.~P.~Y. Th\'{e}bault (2016a).
\newblock {Schr{\"o}dinger Evolution for the Universe: Reparametrization}.
\newblock {\em Classical and Quantum Gravity\/}~{\em 33\/}(6), 065004.

\bibitem[\protect\citeauthoryear{Gryb and Th\'{e}bault}{Gryb and
  Th\'{e}bault}{2016b}]{Gryb:2015}
Gryb, S. and K.~P.~Y. Th\'{e}bault (2016b).
\newblock Time remains.
\newblock {\em British Journal for the Philosophy of Science\/}~{\em 67\/}(3),
  663--705.

\bibitem[\protect\citeauthoryear{Gryb and Th\'{e}bault}{Gryb and
  Th\'{e}bault}{2017}]{Gryb:2017a}
Gryb, S. and K.~P.~Y. Th\'{e}bault (2017).
\newblock Bouncing unitary cosmology i. mini-superspace general solution.
\newblock Preprint.

\bibitem[\protect\citeauthoryear{Gryb and Th\'{e}bault}{Gryb and
  Th\'{e}bault}{2024}]{Gryb:2024}
Gryb, S.~B. and K.~P.~Y. Th\'{e}bault (2024).
\newblock {\em Time Regained: Symmetry and Evolution in Classical Mechanics}.
\newblock Oxford University Press (In Press).

\bibitem[\protect\citeauthoryear{Halliwell and Hawking}{Halliwell and
  Hawking}{1985}]{halliwell:1985}
Halliwell, J.~J. and S.~W. Hawking (1985).
\newblock Origin of structure in the universe.
\newblock {\em Physical Review D\/}~{\em 31\/}(8), 1777.

\bibitem[\protect\citeauthoryear{Hartle}{Hartle}{1993}]{hartle:1993}
Hartle, J.~B. (1993).
\newblock Spacetime quantum mechanics and the quantum mechanics of spacetime.
\newblock {\em arXiv preprint gr-qc/9304006\/}.

\bibitem[\protect\citeauthoryear{Hartle and Hawking}{Hartle and
  Hawking}{1983}]{hartle:1983}
Hartle, J.~B. and S.~W. Hawking (1983).
\newblock Wave function of the universe.
\newblock {\em Physical Review D\/}~{\em 28\/}(12), 2960.

\bibitem[\protect\citeauthoryear{Hawking}{Hawking}{1985}]{Hawking:1985}
Hawking, S.~W. (1985).
\newblock Arrow of time in cosmology.
\newblock {\em Physical Review D\/}~{\em 32\/}(10), 2489--2495.

\bibitem[\protect\citeauthoryear{Hawking}{Hawking}{1994}]{hawking:1994}
Hawking, S.~W. (1994).
\newblock The no boundary condition and the arrow of time.
\newblock In Halliwell, Perez-Mercader, and Zurek (Eds.), {\em Physical Origins
  of Time Asymmetry}, pp.\  346--357. Cambridge University Press.

\bibitem[\protect\citeauthoryear{Hawking, Laflamme, and Lyons}{Hawking
  et~al.}{1993}]{hawking:1993}
Hawking, S.~W., R.~Laflamme, and G.~W. Lyons (1993).
\newblock Origin of time asymmetry.
\newblock {\em Physical Review D\/}~{\em 47\/}(12), 5342.

\bibitem[\protect\citeauthoryear{Hendry}{Hendry}{1998}]{hendry:1998}
Hendry, R.~F. (1998).
\newblock Models and approximations in quantum chemistry.
\newblock {\em Poznan Studies in the Philosophy of the Sciences and the
  Humanities\/}~{\em 63}, 123--142.

\bibitem[\protect\citeauthoryear{Hendry}{Hendry}{2006}]{hendry:2006}
Hendry, R.~F. (2006).
\newblock {\em Is there downward causation in chemistry?}
\newblock Springer.

\bibitem[\protect\citeauthoryear{Hendry}{Hendry}{2010a}]{hendry:2010}
Hendry, R.~F. (2010a).
\newblock Emergence vs. reduction in chemistry.
\newblock {\em Emergence in mind\/}, 205--221.

\bibitem[\protect\citeauthoryear{Hendry}{Hendry}{2010b}]{hendry:2010b}
Hendry, R.~F. (2010b).
\newblock Ontological reduction and molecular structure.
\newblock {\em Studies in History and Philosophy of Science Part B: Studies in
  History and Philosophy of Modern Physics\/}~{\em 41\/}(2), 183--191.

\bibitem[\protect\citeauthoryear{Hendry}{Hendry}{2017}]{hendry:2017}
Hendry, R.~F. (2017).
\newblock Prospects for strong emergence in chemistry.
\newblock In {\em Philosophical and scientific perspectives on downward
  causation}, pp.\  146--163. Routledge.

\bibitem[\protect\citeauthoryear{Henneaux and Teitelboim}{Henneaux and
  Teitelboim}{1992}]{Henneaux:1992a}
Henneaux, M. and C.~Teitelboim (1992).
\newblock {\em Quantization of gauge systems}.
\newblock Princeton University Press.

\bibitem[\protect\citeauthoryear{Hettema}{Hettema}{2017}]{hettema:2017}
Hettema, H. (2017).
\newblock The union of chemistry and physics.
\newblock {\em Cham: Springer International\/}.

\bibitem[\protect\citeauthoryear{Hollowood and McDonald}{Hollowood and
  McDonald}{2017}]{hollowood:2017}
Hollowood, T.~J. and J.~I. McDonald (2017).
\newblock Decoherence, discord, and the quantum master equation for
  cosmological perturbations.
\newblock {\em Physical Review D\/}~{\em 95\/}(10), 103521.

\bibitem[\protect\citeauthoryear{Huggett}{Huggett}{2021}]{huggett2021spacetime}
Huggett, N. (2021).
\newblock Spacetime ``emergence''.
\newblock In {\em The Routledge companion to philosophy of physics}, pp.\
  374--385. Routledge.

\bibitem[\protect\citeauthoryear{Huggett and Wuthrich}{Huggett and
  Wuthrich}{2021}]{huggett2021nowhere}
Huggett, N. and C.~Wuthrich (2021).
\newblock Out of nowhere: Introduction: The emergence of spacetime.

\bibitem[\protect\citeauthoryear{Jecko}{Jecko}{2014}]{Jecko:2014}
Jecko, T. (2014).
\newblock On the mathematical treatment of the born-oppenheimer approximation.
\newblock {\em Journal of Mathematical Physics\/}~{\em 55\/}(5), 053504.

\bibitem[\protect\citeauthoryear{Kiefer}{Kiefer}{1987}]{kiefer:1987}
Kiefer, C. (1987).
\newblock Continuous measurement of mini-superspace variables by higher
  multipoles.
\newblock {\em Classical and Quantum Gravity\/}~{\em 4\/}(5), 1369.

\bibitem[\protect\citeauthoryear{Kiefer}{Kiefer}{1988}]{Kiefer:1988}
Kiefer, C. (1988, Sep).
\newblock Wave packets in minisuperspace.
\newblock {\em Phys. Rev. D\/}~{\em 38\/}(6), 1761--1772.

\bibitem[\protect\citeauthoryear{Kiefer}{Kiefer}{1990}]{kiefer:1990}
Kiefer, C. (1990).
\newblock Wave packets in quantum cosmology and the cosmological constant.
\newblock {\em Nuclear Physics B\/}~{\em 341\/}(1), 273--293.

\bibitem[\protect\citeauthoryear{Kiefer}{Kiefer}{2005}]{kiefer:2005}
Kiefer, C. (2005).
\newblock The semiclassical approximation to quantum gravity.
\newblock In {\em Canonical Gravity: From Classical to Quantum: Proceedings of
  the 117th WE Heraeus Seminar Held at Bad Honnef, Germany, 13--17 September
  1993}, pp.\  170--212. Springer.

\bibitem[\protect\citeauthoryear{Kiefer}{Kiefer}{2009}]{kiefer:2009}
Kiefer, C. (2009).
\newblock Quantum geometrodynamics: whence, whither?
\newblock {\em General Relativity and Gravitation\/}~{\em 41\/}(4), 877--901.

\bibitem[\protect\citeauthoryear{Kiefer}{Kiefer}{2012}]{Kiefer:2012}
Kiefer, C. (2012).
\newblock {\em Quantum Gravity}.
\newblock International Series of Monographs on Physics. Clarendon Press,
  Oxford.

\bibitem[\protect\citeauthoryear{Kiefer}{Kiefer}{2013}]{kiefer:2013}
Kiefer, C. (2013).
\newblock Conceptual problems in quantum gravity and quantum cosmology.
\newblock {\em International Scholarly Research Notices\/}~{\em 2013}.

\bibitem[\protect\citeauthoryear{Kiefer}{Kiefer}{2022}]{kiefer:2022b}
Kiefer, C. (2022).
\newblock On a quantum weyl curvature hypothesis.
\newblock {\em AVS Quantum Science\/}~{\em 4\/}(1), 015607.

\bibitem[\protect\citeauthoryear{Kiefer and Peter}{Kiefer and
  Peter}{2022}]{kiefer:2022}
Kiefer, C. and P.~Peter (2022).
\newblock Time in quantum cosmology.
\newblock {\em Universe\/}~{\em 8\/}(1), 36.

\bibitem[\protect\citeauthoryear{Kiefer and Zeh}{Kiefer and
  Zeh}{1995}]{kiefer:1995}
Kiefer, C. and H.~Zeh (1995).
\newblock Arrow of time in a recollapsing quantum universe.
\newblock {\em Physical Review D\/}~{\em 51\/}(8), 4145.

\bibitem[\protect\citeauthoryear{Komar}{Komar}{1971}]{Komar:1971}
Komar, A. (1971).
\newblock General-relativistic observables via hamilton-jacobi functionals.
\newblock {\em Physical Review D\/}~{\em 4\/}(4), 923--927.

\bibitem[\protect\citeauthoryear{Kucha{\v{r}}}{Kucha{\v{r}}}{2011}]{kuchavr:2011}
Kucha{\v{r}}, K.~V. (2011).
\newblock Time and interpretations of quantum gravity.
\newblock {\em International Journal of Modern Physics D\/}~{\em 20\/}(supp01),
  3--86.

\bibitem[\protect\citeauthoryear{Kucha{\v{r}} and Ryan}{Kucha{\v{r}} and
  Ryan}{1989}]{kuchavr:1989}
Kucha{\v{r}}, K.~V. and M.~P. Ryan (1989).
\newblock Is minisuperspace quantization valid?: Taub in mixmaster.
\newblock {\em Physical Review D\/}~{\em 40\/}(12), 3982.

\bibitem[\protect\citeauthoryear{Ladyman and Th\'{e}bault}{Ladyman and
  Th\'{e}bault}{2024}]{Ladyman:2023}
Ladyman, J. and K.~P.~Y. Th\'{e}bault (2024).
\newblock Whether the universe is a closed system is an open question.
\newblock In M.~E. Cuffaro and S.~Hartmann (Eds.), {\em The Open Systems View:
  Physics, Metaphysics and Methodology}. Oxford University Press.

\bibitem[\protect\citeauthoryear{Maniccia, De~Angelis, and Montani}{Maniccia
  et~al.}{2022}]{maniccia:2022}
Maniccia, G., M.~De~Angelis, and G.~Montani (2022).
\newblock Wkb approaches to restore time in quantum cosmology: Predictions and
  shortcomings.
\newblock {\em Universe\/}~{\em 8\/}(11), 556.

\bibitem[\protect\citeauthoryear{McTaggart}{McTaggart}{1908}]{mctaggart:1908}
McTaggart, J.~E. (1908).
\newblock The unreality of time.
\newblock {\em Mind\/}, 457--474.

\bibitem[\protect\citeauthoryear{Messiah}{Messiah}{1962}]{messiah:1962}
Messiah, A. (1962).
\newblock {\em Quantum mechanics: volume II}.
\newblock North-Holland Publishing Company Amsterdam.

\bibitem[\protect\citeauthoryear{Misner, Thorne, and Wheeler}{Misner
  et~al.}{1974}]{Misner:1974}
Misner, C., K.~Thorne, and J.~Wheeler (1974).
\newblock {\em Gravitation}.
\newblock W.H Freeman and Company.

\bibitem[\protect\citeauthoryear{Newton-Smith}{Newton-Smith}{1982}]{newton:1982}
Newton-Smith, W. (1982).
\newblock {\em The Structure of Time}.
\newblock Routledge \& Kegan Paul.

\bibitem[\protect\citeauthoryear{Palacios}{Palacios}{2022}]{Palacios:2022}
Palacios, P. (2022).
\newblock {\em Emergence and Reduction in Physics}.
\newblock Cambridge University Press.

\bibitem[\protect\citeauthoryear{Peres}{Peres}{1962}]{Peres:1962}
Peres, A. (1962).
\newblock On cauchy's problem in general relativity - ii.
\newblock {\em Il Nuovo Cimento\/}~{\em 26\/}(1), 53--62.

\bibitem[\protect\citeauthoryear{Price}{Price}{1996}]{price:1996}
Price, H. (1996).
\newblock {\em Time's arrow \& Archimedes' point: new directions for the
  physics of time}.
\newblock Oxford University Press, USA.

\bibitem[\protect\citeauthoryear{Price}{Price}{2004}]{price2004origins}
Price, H. (2004).
\newblock {\em On the origins of the arrow of time: Why there is still a puzzle
  about the low-entropy past?}, pp.\  219--239.
\newblock London: Blackwell.

\bibitem[\protect\citeauthoryear{Reichenbach}{Reichenbach}{1956}]{reichenbach:1956}
Reichenbach, H. (1956).
\newblock {\em The Direction of Time}.
\newblock University of California Press.

\bibitem[\protect\citeauthoryear{Robertson}{Robertson}{2020}]{robertson:2020}
Robertson, K. (2020).
\newblock Asymmetry, abstraction, and autonomy: Justifying coarse-graining in
  statistical mechanics.
\newblock {\em The British Journal for the Philosophy of Science\/}.

\bibitem[\protect\citeauthoryear{Ryder}{Ryder}{2022}]{Ryder:2022}
Ryder, D.~J. (2022, December).
\newblock Directed temporal asymmetry from scale invariant dynamics: Is the
  problem of time's arrow solved?
\newblock {\em philsci-archive.pitt.edu/21530\/}.

\bibitem[\protect\citeauthoryear{Sakurai and Commins}{Sakurai and
  Commins}{1995}]{sakurai:1995}
Sakurai, J.~J. and E.~D. Commins (1995).
\newblock {\em Modern quantum mechanics, revised edition}.
\newblock American Association of Physics Teachers.

\bibitem[\protect\citeauthoryear{Salisbury}{Salisbury}{2020}]{Salisbury:2020}
Salisbury, D. (2020).
\newblock Observables and hamilton-jacobi approaches to general relativity.
\newblock {\em Unpublished\/}.

\bibitem[\protect\citeauthoryear{Scerri}{Scerri}{2012}]{scerri:2012}
Scerri, E.~R. (2012).
\newblock Top-down causation regarding the chemistry--physics interface: a
  sceptical view.
\newblock {\em Interface Focus\/}~{\em 2\/}(1), 20--25.

\bibitem[\protect\citeauthoryear{Seifert}{Seifert}{2020}]{seifert:2020}
Seifert, V.~A. (2020).
\newblock The strong emergence of molecular structure.
\newblock {\em European Journal for Philosophy of Science\/}~{\em 10\/}(3), 45.

\bibitem[\protect\citeauthoryear{Seifert}{Seifert}{2022}]{seifert:2022}
Seifert, V.~A. (2022).
\newblock Do molecules have structure in isolation? how models can provide the
  answer.
\newblock In {\em Philosophical Perspectives in Quantum Chemistry}, pp.\
  125--143. Springer.

\bibitem[\protect\citeauthoryear{Sloan}{Sloan}{2021}]{sloan:2021}
Sloan, D. (2021).
\newblock New action for cosmology.
\newblock {\em Physical Review D\/}~{\em 103\/}(4), 043524.

\bibitem[\protect\citeauthoryear{Sloan}{Sloan}{2023}]{sloan:2023}
Sloan, D. (2023).
\newblock Herglotz action for homogeneous cosmologies.
\newblock {\em Classical and Quantum Gravity\/}.

\bibitem[\protect\citeauthoryear{Tarr{\'\i}o, Fern{\'a}ndez-M{\'e}ndez, and
  Marug{\'a}n}{Tarr{\'\i}o et~al.}{2013}]{tarrio:2013}
Tarr{\'\i}o, P., M.~Fern{\'a}ndez-M{\'e}ndez, and G.~A.~M. Marug{\'a}n (2013).
\newblock Singularity avoidance in the hybrid quantization of the gowdy model.
\newblock {\em Physical Review D\/}~{\em 88\/}(8), 084050.

\bibitem[\protect\citeauthoryear{Te~Vrugt, Hossenfelder, and
  Wittkowski}{Te~Vrugt et~al.}{2021}]{te:2021}
Te~Vrugt, M., S.~Hossenfelder, and R.~Wittkowski (2021).
\newblock Mori-zwanzig formalism for general relativity: a new approach to the
  averaging problem.
\newblock {\em Physical Review Letters\/}~{\em 127\/}(23), 231101.

\bibitem[\protect\citeauthoryear{Th{\'e}bault}{Th{\'e}bault}{2023}]{Thebault:2023}
Th{\'e}bault, K.~P. (2023).
\newblock Big bang singularity resolution in quantum cosmology.
\newblock {\em Classical and Quantum Gravity\/}~{\em 40\/}(5), 055007.

\bibitem[\protect\citeauthoryear{Vilenkin}{Vilenkin}{1984}]{vilenkin:1984}
Vilenkin, A. (1984).
\newblock Quantum creation of universes.
\newblock {\em Physical Review D\/}~{\em 30\/}(2), 509.

\bibitem[\protect\citeauthoryear{Vilenkin}{Vilenkin}{1989}]{vilenkin1989interpretation}
Vilenkin, A. (1989).
\newblock Interpretation of the wave function of the universe.
\newblock {\em Physical Review D\/}~{\em 39\/}(4), 1116.

\bibitem[\protect\citeauthoryear{Wallace}{Wallace}{2011}]{Wallace:2011}
Wallace, D. (2011, November).
\newblock The logic of the past hypothesis.
\newblock To appear in B. Loewer, E. Winsberg and B. Weslake (ed.),
  currently-untitled volume discussing David Albert's "Time and Chance".

\bibitem[\protect\citeauthoryear{Warrier}{Warrier}{2022}]{warrier2022case}
Warrier, N. (2022).
\newblock The case of the vanishing wavefunction.
\newblock {\em Studies in History and Philosophy of Science\/}~{\em 96},
  135--140.

\bibitem[\protect\citeauthoryear{Weatherall}{Weatherall}{2018}]{weatherall:2018}
Weatherall, J.~O. (2018).
\newblock Regarding the `hole argument'.
\newblock {\em The British Journal for the Philosophy of Science\/}~{\em
  69\/}(2), 329--350.

\bibitem[\protect\citeauthoryear{Wheeler}{Wheeler}{1968}]{Wheeler:1968}
Wheeler, J.~A. (1968).
\newblock {Superspace and the Nature of Quantum Geometrodynamics}.
\newblock In J.~A. Wheeler and C.~M. De~Witt (Eds.), {\em Lectures in
  Mathematics and Physics}. Benjamin.

\bibitem[\protect\citeauthoryear{Woolley and Sutcliffe}{Woolley and
  Sutcliffe}{1977}]{woolley:1977}
Woolley, R. and B.~Sutcliffe (1977).
\newblock Molecular structure and the born---oppenheimer approximation.
\newblock {\em Chemical Physics Letters\/}~{\em 45\/}(2), 393--398.

\bibitem[\protect\citeauthoryear{Woolley}{Woolley}{1978}]{woolley:1978}
Woolley, R.~G. (1978).
\newblock Must a molecule have a shape?
\newblock {\em Journal of the American Chemical Society\/}~{\em 100\/}(4),
  1073--1078.

\bibitem[\protect\citeauthoryear{Zeh}{Zeh}{1986}]{zeh:1986}
Zeh, H. (1986).
\newblock Emergence of classical time from a universal wavefunction.
\newblock {\em Physics Letters A\/}~{\em 116\/}(1), 9--12.

\bibitem[\protect\citeauthoryear{Zeh}{Zeh}{1989}]{zeh:1989}
Zeh, H.-D. (1989).
\newblock {\em The direction of time}.
\newblock Springer.

\end{thebibliography}

\bigskip

\begin{center}
    \textbf{Appendix}
\end{center}

\begin{appendix}

In this Appendix we sketch proofs of formal claims made in the discussion of the BO approximation in \S\ref{WDWordmetric}.\\

\noindent [1] Given $|\lambda_m-\lambda_n|\gg T_1$ for $m>n$, a superposition of $\psi_n$ cannot be an eigenvector of total energy, so \eqref{eq:BO2} holds. Suppose, for reductio, that (for instance) $\theta_m\psi_m+\theta_n\psi_n$ is an eigenstate, so that it is parallel with $(\hat T_1+\hat T_2+\hat W\big)(\theta_m\psi_m+\theta_n\psi_n) = \hat T_1(\theta_m\psi_m+\theta_n\psi_n) + \lambda_m\theta_m\psi_m+\lambda_n\theta_n\psi_n$. Since the $\psi_n$ are orthonormal, such parallelism could only hold if the first term is comparable to the difference between the second two: that the kinetic energy of the heavy subsystem is comparable to the difference between the $\lambda_n$, contrary to the separation of energy levels.\\

\noindent [2] The adiabatic approximation, \eqref{eq:BO3}: $\theta_n\psi_n$ is an eigenstate of $\hat T_2+\hat W$ from \eqref{eq:BO1}, and (approximately) of $\hat T_1+\hat T_2+\hat W$ from \eqref{eq:BO2}, hence it is also an eigenstate of $\hat T_1$. So by the orthogonality of the $\psi_n$, $\langle\theta_m\psi_m|\hat T_1|\theta_n\psi_n\rangle\propto\delta_{m,n}$. In the $x_i$ basis $\hat T_1\sim\partial^2/\partial x_1^2$, so we have (using the chain rule):

\begin{equation}
    \nonumber\int\mathrm{d}x_1\mathrm{d}x_2\ \theta^*_m\psi^*_m\big(\frac{\partial^2\theta_n}{\partial x_1^2}\psi_n + 2\frac{\partial\theta_n}{\partial x_1}\frac{\partial\psi_n}{\partial x_1} + \theta_n\frac{\partial^2\psi_n}{\partial x_1^2}\big) \propto \delta_{m,n}.
\end{equation}
The orthogonality of the $\psi_n$ means that (the integral of) the first term is proportional to $\delta_{m,n}$, while the second and third terms are generally not. Hence the (approximate) proportionality requires that those terms -- specifically the $\psi_n$ derivatives that they contain -- (approximately) vanish. Hence \eqref{eq:BO3}.
\end{appendix}

\end{document}